\documentclass[twocolumn,traditabstract]{aa}

\usepackage[english]{babel}
\usepackage{graphicx,amsmath}
\usepackage{epstopdf}
\usepackage{epsf,color}
\usepackage[mathscr]{eucal}
\usepackage{amsmath}
\usepackage{amssymb,amsfonts}
\usepackage{natbib}
\bibpunct{(}{)}{;}{a}{}{,} 

\newcommand{\master}{{\small M{\tiny ASTER}}}
\newcommand{\rv}{{\vec{r}}}
\newcommand{\kv}{{\vec{k}}}
\newcommand{\poker}{{\small P{\tiny OKER}}}

\def\simlt{\lower.5ex\hbox{$\; \buildrel < \over \sim \;$}}
\def\simgt{\lower.5ex\hbox{$\; \buildrel > \over \sim \;$}}

\title{POKER: Estimating the power spectrum of diffuse emission with complex masks and
  at high angular resolution}
\author{N. Ponthieu, J. Grain, G. Lagache}
\offprints{N. Ponthieu, \email{Nicolas.Ponthieu@ias.u-psud.fr}}
\institute{IAS, Institut d'Astrophysique Spatiale, CNRS Universit\'e Paris 11, B\^atiment 121, 91405 Orsay, France}

\titlerunning{P. Of $k$ EstimatoR}
\authorrunning{Ponthieu et al.}

\date{\today}

\abstract{We describe the implementation of an angular power spectrum estimator
  in the flat sky approximation. \poker~(P. Of $k$ EstimatoR) is based on the
 \master~algorithm developped by Hivon and collaborators in the context of CMB
  anisotropy. It works entirely in discrete space and can be applied to
  arbitrary high angular resolution maps. It is therefore particularly suitable
  for current and future infrared to sub-mm observations of diffuse emission,
  whether Galactic or cosmological.}

\begin{document}
\maketitle

\section{Introduction}

Whether it is due to Galactic dust or synchrotron, to cosmological
backgrounds such as the Cosmic Microwave Background (CMB) or to the Cosmic
Infrared Background (CIB), that traces the integrated radiation of unresolved
galaxies, diffuse emission is omnipresent in infrared and millimetric
observations. The angular power spectrum of this radiation is one of the main
tools used to constrain the structure of the interstellar medium, the clustering
of IR galaxies (CIB), or the cosmological parameters (CMB). In short, its
estimation requires to Fourier transform the image and to average the modulus
square of the Fourier amplitudes into frequency bins. However, the image has
both boundaries and often masked regions (e.g. to remove bright point sources)
that induce power aliasing and biases the estimation of the power spectrum if
not accounted for properly. The effect becomes quite significant when the signal
has a steep power spectrum such as $k^{-3}$, similar to that measured for
Galactic cirrus emission \citep{mamd_07} or even steeper than $k^{-4}$ as for
CMB anisotropy on angular scales smaller than a few arcmin
\citep[e.g.][]{acbar,quad_t}. To account for non-periodic boundaries, \cite{das}
proposed an original apodizing technique that helps us to deconvolve the
estimated power spectrum from that of the observed patch boundaries. These
authors also mitigate the impact of holes by applying a pre-whitening technique
to data in real space.

In the context of CMB anisotropy, \cite{master} developed the \master~method
that allows us to correct for mask effects on the output binned power
spectrum. They analyze data accross the full sky and account for the sky
curvature. Instead of classical Fourier analysis, they project the data onto
spherical harmonics and go through the algebra of \emph{pseudo} angular power
spectra (see Sect.~\ref{se:ps_def}). This idea has been successfully used in
several experiments \citep[e.g.][]{boomerang,archeops_t1} and is also the basis
of more refined algorithms used in e.g.~WMAP \citep{hinshaw} and Archeops
\citep{archeops_t2}. However, direct use of \master~in the context of infrared
observations with a resolution of typically a few arcsec requires us to estimate
Legendre polynomials up to orders $\ell$ of 10,000 or more for which current
recurrences and integration methods are numerically unstable. Other techniques
developed in the context of CMB anisotropy such as maximum likelihood estimation
\citep{bjk} could be transposed to high angular resolution maps, but the
numerical cost $\propto n_{pix}^3$ is prohibitive for common applications when
the analysis pipeline requires Monte Carlo simulations.

This paper aims to transpose the \emph{pseudo}-spectrum approach pioneered by
\master~ to high angular resolution observations and a classical Fourier
analysis in the context of the flat sky approximation. Its originality compared
to other approaches is that it works exclusively in discrete space and therefore
avoids the complexity of resampling the data and integrating Bessel
functions. Our algorithm was nicknamed \poker, for ``P. Of $k$ EstimatoR''. The
paper is organized as follows. Section~\ref{se:ps_def} provides the definitions
and algebra involved in \poker~and Sect.~\ref{se:applis} shows its applications
to simulations of various astrophysical components spectra and a complex
mask. Detailed derivations of our results are presented in the appendices.
Although this work focuses on temperature power spectrum estimation, we also
show in Appendices \ref{se:app_eb} and \ref{se:app_te} how the formalism can be
generalized to the case of polarization.

\section{Power spectrum estimation for an incomplete observation of the sky}
\label{se:ps_def}

We first briefly state the limits of the flat sky approximation, then recall
the definitions of both the power spectrum and \emph{pseudo}-power spectrum of data in the
context of continuous Fourier transforms. Finally, we consider their counterpart
in discrete space and the implementation of \poker.

\subsection{Flat sky approximation}

Projecting an observed fraction of the sky onto a plane rather than a sphere
alters the image properties in a way that depends on the specific reprojection
scheme (e.g. gnomonic, tangential, cylindrical etc). The angular power spectrum
of the data as measured on the projection plane therefore differs from that on
the sphere. Two comments can be made at this stage. First, as long as the
observation patch does not span more than a few degrees, as in most
infrared experiments, the distortions are very small. For a gnomonic
projection for instance, an observation point at an angular distance $\theta$
from the map center (the point tangent to the sphere) is projected at a distance
$\tan(\theta)$ rather than $\theta$. The relative difference between the two is
only $2.6\;10^{-3}$ for a map of 10 degree diameter i.e. the projected map
  is stretched by 46~arcsec in each direction. As long as this remains small
  compared to the angular scales over which the angular power spectrum is
  estimated, the distortion can be neglected \citep[e.g.][]{pryke}. In
  this work, we stay well above this limit by considering square maps of only
  $5\times 5$~deg$^2$ and 2~arcmin resolution and therefore neglect distortion
  effects. Second, the impact of the projection on the estimation of the power
spectrum is equivalent to a transfer function that can be estimated and
corrected for when dealing with the convolution kernel (Eq.~\ref{eq:ps2cl_1}),
as detailed in Sect.~\ref{se:transf_func}. Both of these reasons allow us to
estimate the angular power spectrum of a map built in the flat sky approximation
limit.

\subsection{Continuous Fourier analysis and masked data}

On a flat two-dimensional (2D) surface, a scalar field $T_\rv$ depending on the
direction of observation $\rv$ is represented in Fourier space by

\begin{eqnarray}
T_\kv & = & \int_{\mathbb{R}^2} d\rv T_\rv e^{-i\kv\cdot\rv}, \label{eq:ft_def1} \\
T_\rv & = & \int_{\mathbb{R}^2}\frac{d\kv}{(2\pi)^2}   T_\kv
e^{i\kv\cdot\rv}. \label{eq:ft_def2}
\end{eqnarray}

\noindent For a random isotropic process, the 2D power spectrum $\mathcal{P}_\kv$ is defined as

\begin{equation}
\langle T_\kv T^*_{\kv^\prime}\rangle \equiv \mathcal{P}_\kv\delta_{\kv-\kv^\prime},
\label{eq:cont_2dps_def}
\end{equation}

\noindent where the brackets denote the statistical average. We denote by
one-dimensional (1D) power spectrum its azimuthal average

\begin{equation}
P_k \equiv \frac{1}{2\pi}\int_0^{2\pi} d\theta\; T_\kv T^*_\kv\;,
\label{eq:1d_ps_def}
\end{equation}

\noindent where $P_k$ is the physical quantity of interest which we want to reconstruct. It is
the Fourier transform of the two-point correlation function. If the process is
isotropic, the 2D and 1D-power spectra are related by

\begin{equation}
\langle T_\kv T^*_{\kv^\prime}\rangle \equiv \mathcal{P}_\kv
\delta_{\kv-\kv^\prime} = (2\pi)^2 P_k \delta_{\kv-\kv^\prime},
\label{eq:ps_def}
\end{equation}

In the following, we neglect the 1D or 2D qualifiers to improve readability and
in both 1D and 2D cases use the term power spectrum unless the difference needs
to be emphasized. In practice however, the integrals of Eqs.~(\ref{eq:ft_def1},
\ref{eq:ft_def2}) cannot run up to infinity simply because of the limited size
of the observation patch. This is accounted for by a weight function $W_\rv$
applied to the data. Its most simple form is unitary on the data, zero outside
the observation range or where strong sources are masked out. More subtle
choices such as inverse noise variance weighting or apodization
(cf.~Sect.~\ref{se:applis}) are usually used. Instead of the true Fourier
amplitudes, we are then bound to measure the amplitudes of the masked data,
a.k.a. the \emph{pseudo}-amplitudes

\begin{equation}
\hat{T}_\kv = \int_{\mathbb{R}^2} d\rv T_\rv W_\rv e^{-i\kv\cdot\rv} \label{eq:ft_mask}.
\end{equation}

\noindent Equation~(\ref{eq:1d_ps_def}) applied to the \emph{pseudo}-amplitudes gives the
1D-\emph{pseudo}-power spectrum

\begin{equation}
\hat{P}_k = \int_0^\infty k_1dk_1\;K_{kk_1}P_{k_1},
\label{eq:pps}
\end{equation}

\noindent where $K_{kk_1}$ is the mixing matrix that depends on the weighting function
$W_\rv$. To determine the signal power spectrum, we need to solve this equation
for $P_{k_1}$. A detailed derivation of the analytic solution can be found in
\cite{master}. The impact of the instrumental beam, the pixel window
  function, the projection algorithm, the scanning, and the data processing can
  be accounted for in this formalism as transfer functions and incorporated in the
  definition of $K_{kk_1}$ (Sect.~\ref{se:transf_func}). In the next two
  subsections, we focus on mask effects and the global picture of our
  algorithm.

\subsection{Discrete Fourier analysis and \poker}

Any data set is by construction discretely sampled. Computing the quantities
defined in the previous section requires mathematical interpolation and/or
resampling of these data and appropriate integration tools, especially if
the underlying data power spectrum is steep as for Galactic cirrus for which
$P(k)\propto k^{-3}$ \citep{mamd_07}. Rather than dealing with these difficulties,
we keep the native pixelized description of the data and work completely in
discrete space. We use the discrete fourier transform (hereafter DFT) as
provided by data analysis software. For a map of scalar quantity $D_{\mu \nu}$
and size $N_x\times N_y$ pixels, it is defined as

\begin{eqnarray}
D_{mn} & = & \frac{1}{N_xN_y}\sum_{\mu,\nu} D_{\mu\nu}  e^{-2i\pi(\mu m/N_x+\nu n/N_y)}, \label{eq:dft}\\
D_{\mu\nu} & = & \sum_{m,n} D_{mn} e^{+2i\pi(\mu m/N_x+\nu n/N_y)}.
\end{eqnarray}

Throughout this work, although we denote quantities in direct and Fourier
space by the same name, Greek indices denote pixel indices in real space
whereas roman indices refer to amplitudes in Fourier space. Unless stated
otherwise, sums over $\mu$ and $m$ (resp. $\nu$ and $n$) run from 0 to $N_x-1$
(resp. $N_y-1$), $\Delta\theta$ is the angular resolution of the map in radians. For a
given wave-vector $\kv_{mn}$, labeled by the $m$ and $n$ indices, its
corresponding norm is denoted by $k_{mn} =
(2\pi/\Delta\theta)\sqrt{(m^\prime/N_x)^2+(n^\prime/N_y)^2}$ with $m^\prime=m$
(resp. $n^\prime$) if $m\leq N_x/2$ and $m^\prime=N_x-m$ if $m>N_x/2$. This
convention ensures that on small angular scales $k$ matches the multipole $\ell$
used in the description of CMB anisotropy. The Nyquist mode is
$\pi/\Delta\theta$.

\begin{figure}
\begin{center}
\includegraphics[clip, angle=0, scale = 0.4]{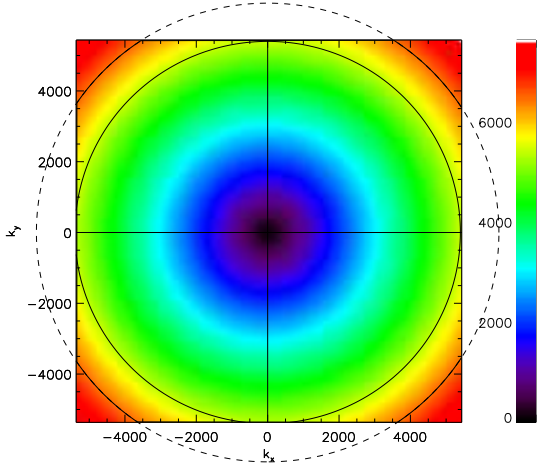}
\caption{Map of the Fourier modes of the worked examples of Sect~\ref{se:applis}. The inner circle
  delimits the Nyquist range. Modes that lie on the outer circle are examples of modes of larger
  modulus than $k_{Nyquist}$. For these modes, not all directions are sampled in
  the Fourier plane (dashes represent the missing modes).}
\label{fig:nyquist}
\end{center}
\end{figure}

It is well known that the DFT slightly differs from the theoretical continuous
Fourier transform, hence $D_{mn}$ does not strictly equal $T_{\kv_{mn}}$. In
particular, the DFT deals with amplitudes for modes $\kv_{mn}$ larger than the
Nyquist mode $\pi/\Delta\theta$ and in some directions for $\theta_{mn}$ only
(see Fig.~\ref{fig:nyquist}). It is therefore not possible to integrate
Eq.~(\ref{eq:1d_ps_def}) on the full range $\theta \in [0,2\pi]$ for such modes
and so, the 1D power spectrum is undefined outside the Nyquist range. In the
following, we therefore restrict ourselves to the Nyquist range for power spectrum
estimation. We note, however, that mathematical sums implied in the following may
still run over the full range of pixels or DFT amplitude indices.

The direct DFT of the masked data results from the convolution of the DFT
amplitudes by a kernel $K_{m,m'}^{n,n'}$ that depends only on the mask DFT
amplitudes (Appendix \ref{se:fsl_dft}). If the data $D$
consist of signal $T$ and noise $N$, we have

\begin{equation}
\langle |\hat{D}_{mn}|^2 \rangle =
\sum_{m'n'} |K_{m,m'}^{n,n'}|^2 |T_{m'n'}|^2 + \langle\hat{N}_{mn}\rangle\;,
\label{eq:ps2cl}
\end{equation}

\noindent which is the transcription in discrete space of Eq.~(\ref{eq:pps}).

The rapid oscillations of the convolution kernel introduce strong correlations
between spatial frequencies and make its inversion numerically
  intractable. (\emph{Pseudo}-)Power spectra are therefore estimated on some
frequency band-powers (labeled $b$ hereafter). The binning operator reads

\begin{equation}
R_b^{mn} = \left\{ \begin{array}{ll}
\frac{k_{mn}^\beta}{\Xi_b}&\;{\rm if}\; k^b_{low} \leq k_{mn} < k^{b+1}_{low}\\
0&\;{\rm otherwise}\end{array}\right. ,
\label{eq:pmat}
\end{equation}

\noindent where $k^b_{low}$ is the mode of lowest modulus that belongs to bin $b$ and $\Xi_b$
is the number of wave vectors $\kv_{mn}$ that fall into
this bin. The reciprocal operator that relates the theoretical value of the
1D binned power spectrum $P_b$ to its value at $\kv_{mn}$ is

\begin{equation}
Q_{mn}^b = \left\{ \begin{array}{ll}
\frac{1}{k_{mn}^\beta}&\;{\rm if}\; k^b_{low} \leq k_{mn} < k^{b+1}_{low}\\
0&\;{\rm otherwise}\end{array}\right. .
\label{eq:qmat}
\end{equation}

Although not strictly required, results may be improved when the spectral
index $\beta$ is chosen so that $k^\beta P_k$ is as flat as possible\footnote{In
the case of CMB, $\beta \simeq 2$ is the equivalent of the standard
$\ell(\ell+1)$ prefactor that flattens the spectrum up to $\ell \sim 2000$.}. In
the case of the cosmic infrared background anisotropy, $\beta \simeq 1$
\citep{planck_cib_2011}. The binned \emph{pseudo}-power spectrum is therefore
given by

\begin{equation}
\hat{P}_b=\sum_{m,n\in b}R^{mn}_b|\hat{T}_{mn}|^2,
\label{eq:r}
\end{equation}

\noindent and the data power spectrum is related to its binned value $P_b$ via

\begin{equation}
|T_{m'n'}|^2 \simeq Q_{m'n'}^{b'}P_{b'}.
\label{eq:q}
\end{equation}

\noindent With such binned quantities, Eq. (\ref{eq:ps2cl}) reads

\begin{equation}
\langle\hat{P}_b\rangle \simeq \sum_{b'} M_{bb'} P_{b'} + \langle
\hat{N}_{b}\rangle,
\label{eq:mbb}
\end{equation}

\noindent where

\begin{equation}
M_{bb'} = \sum_{m,n\in b}\sum_{m',n'\in b'}R_b^{mn} |K_{m,m'}^{n,n'}|^2 Q_{m'n'}^{b'}.
\label{eq:mbb_def}
\end{equation}

\noindent An unbiased estimate of the binned angular power spectrum of the signal is thus given by

\begin{equation}
\tilde{P}_b \simeq \sum_{b'}M^{-1}_{bb'}\left( \hat{P}_{b'} -\langle\hat{N}_{b'}\rangle\right).
\label{eq:pb_res}
\end{equation}

\noindent It can indeed be easily verified that $\langle \tilde{P}_b \rangle =
P_b$. Uncertainties in $\tilde{P}_b$ come from sampling and noise variance that are
estimated via Monte Carlo simulations as described in the next section.

\subsection{Statistical uncertainties \label{sec:stat}}

\begin{figure}
\begin{center}
\includegraphics[clip, angle=0, scale = 0.4]{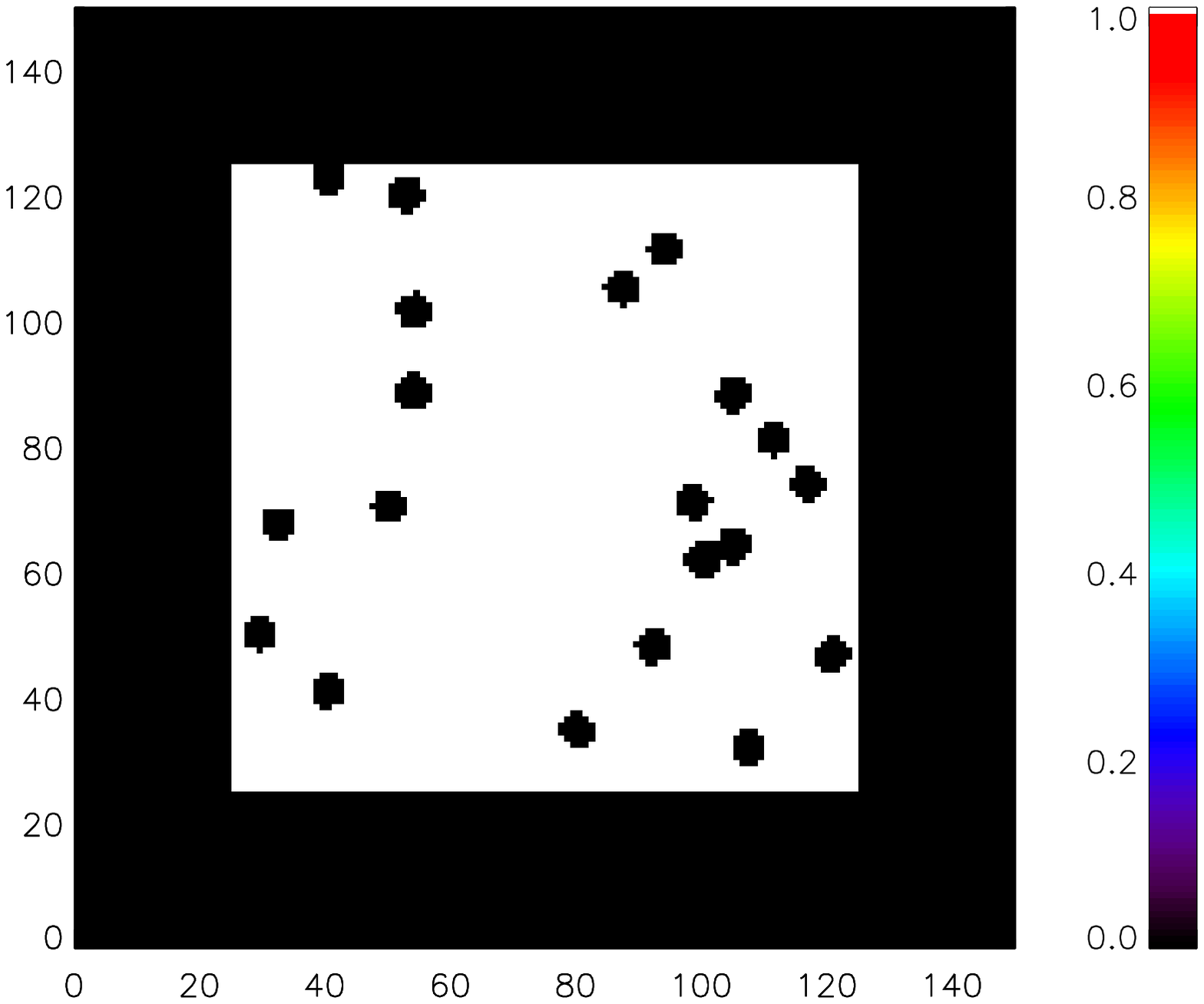}
\includegraphics[clip, angle=0, scale = 0.4]{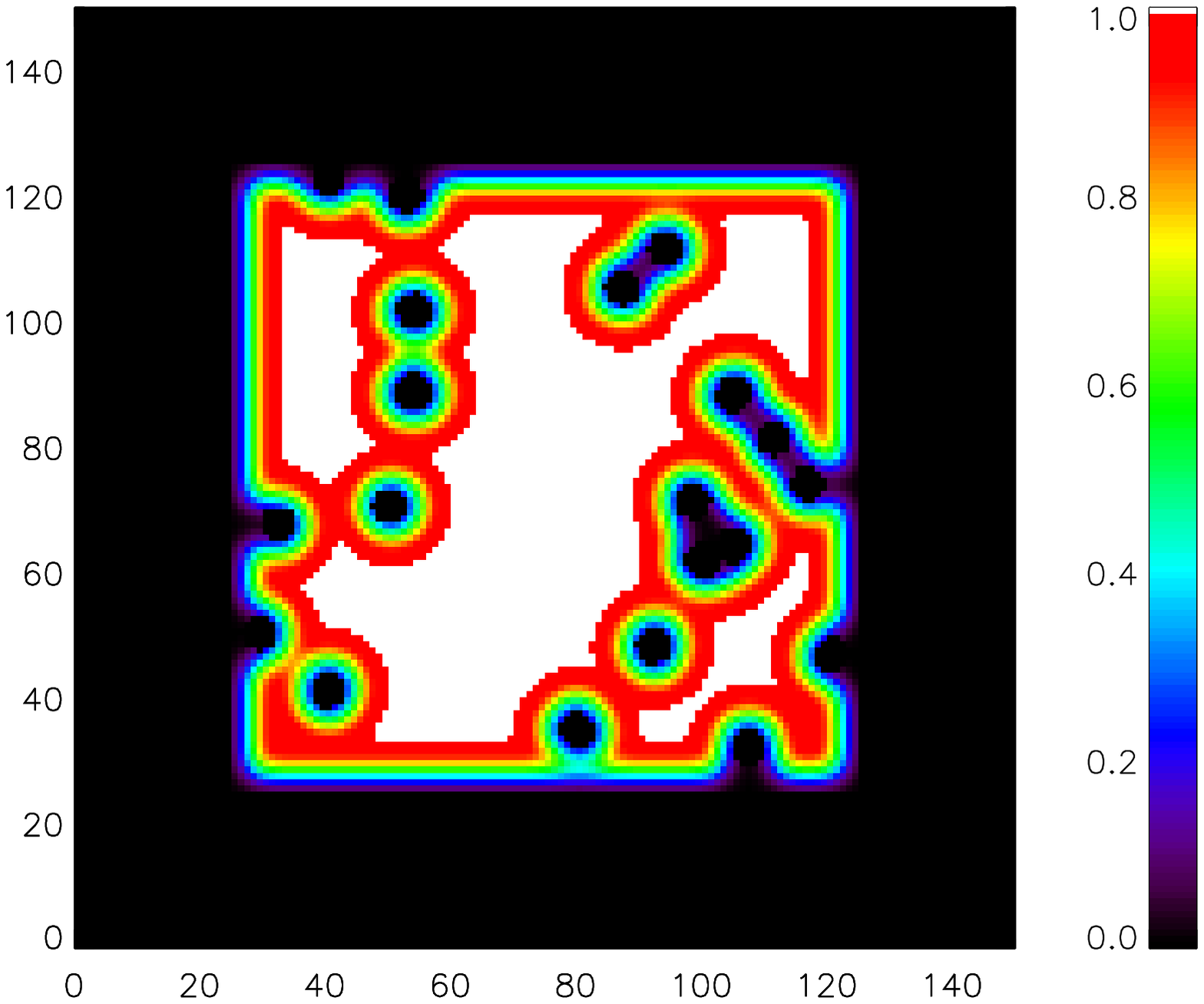}
\caption{\emph{Top:} Mask applied to the simulated data. This mask is 1 where
  data are available and 0 outside the observation patch and where bright sources
  have been masked out. \emph{Bottom:} Same mask with apodized
  boundaries but still with the same number of masked pixels.}
\label{fig:masks}
\end{center}
\end{figure}

Statistical uncertainties in $P_b$ come from signal sampling variance and noise
variance. They are estimated via Monte Carlo simulations. For each realization,
a map of signal+noise is produced (see Sect.~\ref{se:algo}) and treated in
the same way as described for the data in the previous section to give an
estimate, $\tilde{P}_b$, with the same statistical properties as that of the true
data. Altogether, these simulations provide the uncertainties in our
estimate. The covariance matrix of $\tilde{P}_b$ is

\begin{equation}
 {\bf C}_{bb'} = \big< \left(\tilde{P}_b - \langle \tilde{P}_b\rangle_{\text{MC}}\right)\left(\tilde{P}_{b'} -
\langle \tilde{P}_{b'}\rangle_{\text{MC}}\right)\big>_{\text{MC}},
\label{eq:cov_mat}
\end{equation}

\noindent where $\langle\cdot\rangle_{\text{MC}}$ denotes the Monte-Carlo
averaging. The error bar in each $\tilde{P}_b$ is

\begin{equation}
\sigma_{\tilde{P}_b} = \sqrt{{\bf C}_{bb}},
\label{eq:sigma_pb}
\end{equation}

\noindent and the bin-bin correlation matrix is given by its standard definition

\begin{equation}
{\bf \Xi}_{bb'}=\frac{\mathbf{C}_{bb'}}{\sqrt{{\bf C}_{bb}{\bf C}_{b'b'}}}.
\label{eq:xcorr}
\end{equation}

\subsection{Beam, map making and transfer functions \label{se:transf_func}}

The above sections describe the main features of the algorithm that provides an
unbiased estimate of the power spectrum of data \emph{projected onto a
  map}. This power spectrum may however differ from the \emph{signal} power
spectrum. The map making process may indeed alter the statistical properties of
the signal, together with data filtering and the convolution by the instrumental
beam. For instance, the pixelization caused by the map making is equivalent to a
convolution in real space by a square kernel and therefore translates into a
multiplication in Fourier space by a factor $\sin_c$. In the case of CMB
anisotropy for which power spectrum estimation has been most extensively
studied, to a good approximation, the transfer function $F_k$ of the map making
and the data processing together reduces to a function of $k$ that multiplies
the data power spectrum $P_k$. The determination of $F_k$ is performed by a set of
Monte Carlo simulations of data processing and map making. The beam smearing
effect is also described by a multiplicative function $B_k$. In the present
framework, it is possible to be even more precise and to account for the exact
beam shape and orientation because the beam can be completely described by its
Fourier coefficients $B_{mn}$ rather than its approximated annular average
$B_k$. This may be of particular relevance for small fields over which the
scanning directions are approximately constant and increase the
effect of beam asymmetry. The map making together with the filtering transfer
function is also likely to be more accurately represented by a function of both
Fourier indices $F_{mn}$, such that Eq.~(\ref{eq:ps2cl}) is given by

\begin{eqnarray}
\langle |\hat{D}_{mn}|^2 \rangle =&&  \sum_{m'n'} |K_{m,m'}^{n,n'}|^2
F_{m'n'}B_{m'n'}P_{\kv_{m'n'}} \nonumber \\
&& + \langle\hat{N}_{mn}\rangle.
\label{eq:ps2cl_1}
\end{eqnarray}

\noindent These new contributions can be incorporated in the definition of the convolution
kernel $K_{m,m'}^{n,n'}$ such that no additional modification of the algorithm is
needed from Eq.~(\ref{eq:ps2cl}) onward.

\subsection{Algorithm \label{se:algo}}

\begin{figure}
\begin{center}
\includegraphics[clip, angle=0, scale = 0.5]{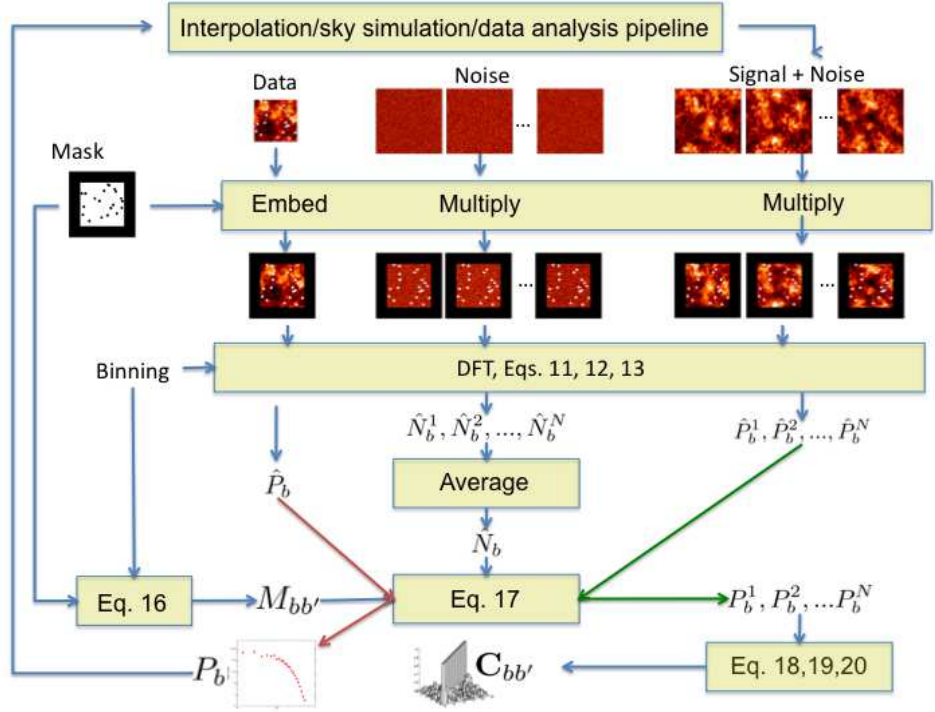}
\caption{Schematic flow chart of \poker.}
\label{fig:chart}
\end{center}
\end{figure}

An outline of the algorithm is presented in Fig.~\ref{fig:chart}. To perform the
complete process of power spectrum and statistical uncertainty estimation, one
needs:

\renewcommand{\labelenumi}{(\alph{enumi})}

\begin{enumerate}
\item A tool to simulate the sky that is observed by the instrument given
  an input angular power spectrum.
\item A tool to simulate the instrument observations given the simulated sky.
\item The data processing pipeline that derives from these observations the map
  from which the user can then estimate the angular power spectrum. The
  pipeline includes optical beam smearing, time domain filtering,
  data flagging, map making etc. This tool is required to determine the transfer function of
  the map making and data processing (Eq.~\ref{eq:ps2cl_1}).
\item A tool to compute the power spectrum of a 2D map and to bin it into a
  set of predefined bins with a weight that may be a function of $k$.
\item A tool to compute $M_{bb'}$, which involves the computation of the
  convolution kernel $K_{m,m'}^{n,n'}$. This is actually the longest part of
  the process because it is a $N_{pix}^2$ operation, but it only needs to be done
  once.
\end{enumerate}

All these tools are provided in the \poker~library\footnote{{\tt
    http://www.ias.u-psud.fr/poker}. This library makes use of some of the HEALPix
    programs \citep{healpix}.}\footnote{The sky simulator provided in the
    library works directly in flat space. Indeed, we cannot anticipate neither
    on the user's favorite reprojection scheme from the sphere to the plane nor
    on his/her data processing pipeline and map making. Furthermore, simulating
    directly a flat sky of a known power spectrum is a convenient first stage to
    optimize the binning and apodization scheme before running the full fledge
    Monte Carlo simulations.}. The algorithm can be summarized as follows:

\renewcommand{\labelenumi}{\arabic{enumi}.}
\begin{enumerate}
\item
Insert the observed sky patch of size $N_x \times N_y$ pixels into a
  ``large patch'' $(N_x^\prime \times N_y^\prime)$ and padd it with zeros. This
  will allow for the correction of aliasing by scales larger than the observed
  sky. The size of the patch and the zero padding that should be used have both to
  be determined by the user. A factor from 1.2 to 2 is enough in most cases. A
  compromise must be chosen between the uncertainty on large scales that the
  user tries to estimate and the uncertainty associated with the unknown power in
  these large scales that needs to be assumed for the simulations. It is also
  possible to apodize the observation patch to limit large-scale aliasing (see
  Sect.~\ref{se:applis} for more details) and improve the bin to bin decorrelation at high $k$.
\item
Define a binning for the estimated power spectrum on the large
  patch. Typically, modes sampled by the data set are the DC level and modes between
  $k_{min} = 2\pi/\Delta\theta/\max(N_x,N_y)$ and the Nyquist mode $k_c =
  \pi/\Delta\theta$. The minimum bandwidth of the bins may be chosen as $\sim 2k_{min}$.
\item
Determine the noise \emph{pseudo}-power spectrum -- $\langle
  \hat{N}_b\rangle$ of Eq.~(\ref{eq:mbb}). If it cannot be determined
  analytically, perform a set of Monte Carlo
  realizations of noise-only maps (with (a)) and compute the power spectrum with
  (b) of the masked maps inserted into the large patch. The average of these
    Monte Carlo realizations gives $\langle\hat{N}_b\rangle$.
\item Run a set of noise-free simulations of the observed and reprojected sky to determine the
  transfer function $F_{mn}$ of the data processing pipeline -- Eq.~(\ref{eq:ps2cl_1}).
\item
Compute $M_{bb'}$ with (c) -- Eqs.~(\ref{eq:pmat}, \ref{eq:qmat}, \ref{eq:mbb_def}). This
  operation scales as $N_p^2$ but it only needs to be done once. The
  implementation proposed in the \poker~library can be run on a multiprocessor
  machine.
\item
Compute the \emph{pseudo}-power spectrum of the masked data on the large
  patch $\hat{P}_b$ with (b) -- Eq.~(\ref{eq:raw_ft}).
\item
Apply Eq.~(\ref{eq:pb_res}) to obtain the binned power spectrum of the data
$P_b$. The resolution of this equation can be done with any suitable method
  of linear algebra. Note that $M_{bb'}$ can be rather small and its inversion
  straightforward with standard numerical tools such that
  Eq.~(\ref{eq:pb_res}) can be computed as is. At this stage, it may be useful
  to discard the first bin of the matrix, which describes the coupling of
  the DC level of the map to the mask and is therefore irrelevant for a power spectrum
  analysis but tends to alter the conditioning of $M_{bb'}$.

\item
Determine the statistical error bars associated with this estimate. For that, perform
a set of Monte Carlo realizations of signal+ noise. The input spectrum required
for these simulations can be a smooth interpolation of the binned power spectrum
determined at the previous step. For each realization, compute the
\emph{pseudo}-power spectrum (using (b) on the masked data embedded in the large
patch), subtract $\langle \hat{N}_b\rangle$, and then solve Eq.~(\ref{eq:pb_res}). This
provides a set of random realizations of $\tilde{P}_b$. The error bars and the
bin-to-bin covariance matrix are then given by
Eqs.~(\ref{eq:cov_mat},\ref{eq:sigma_pb}).
\end{enumerate}

\begin{figure}
\begin{center}
\includegraphics[clip, angle=0, scale = 0.4]{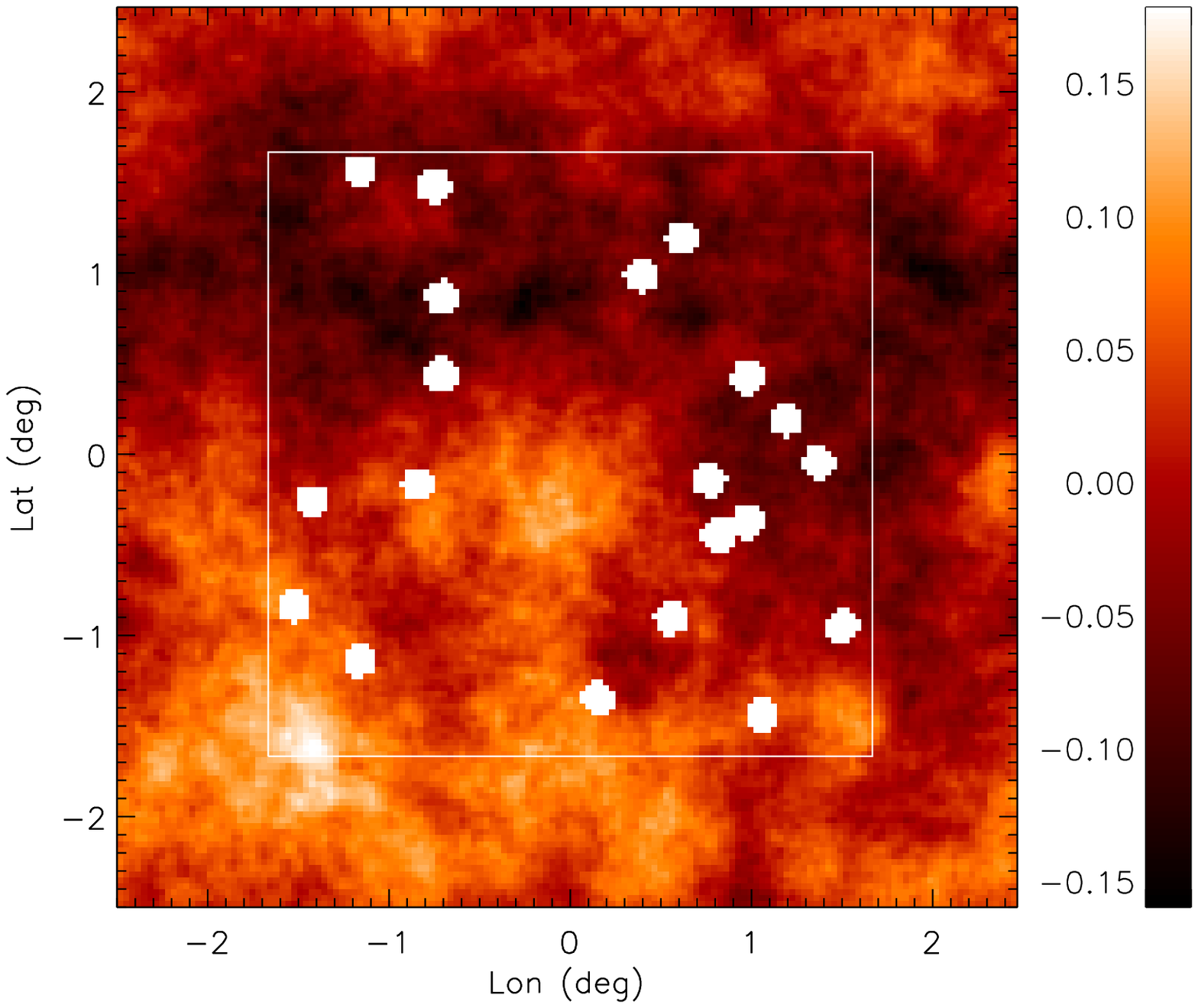}
\includegraphics[clip, angle=0, scale = 0.4]{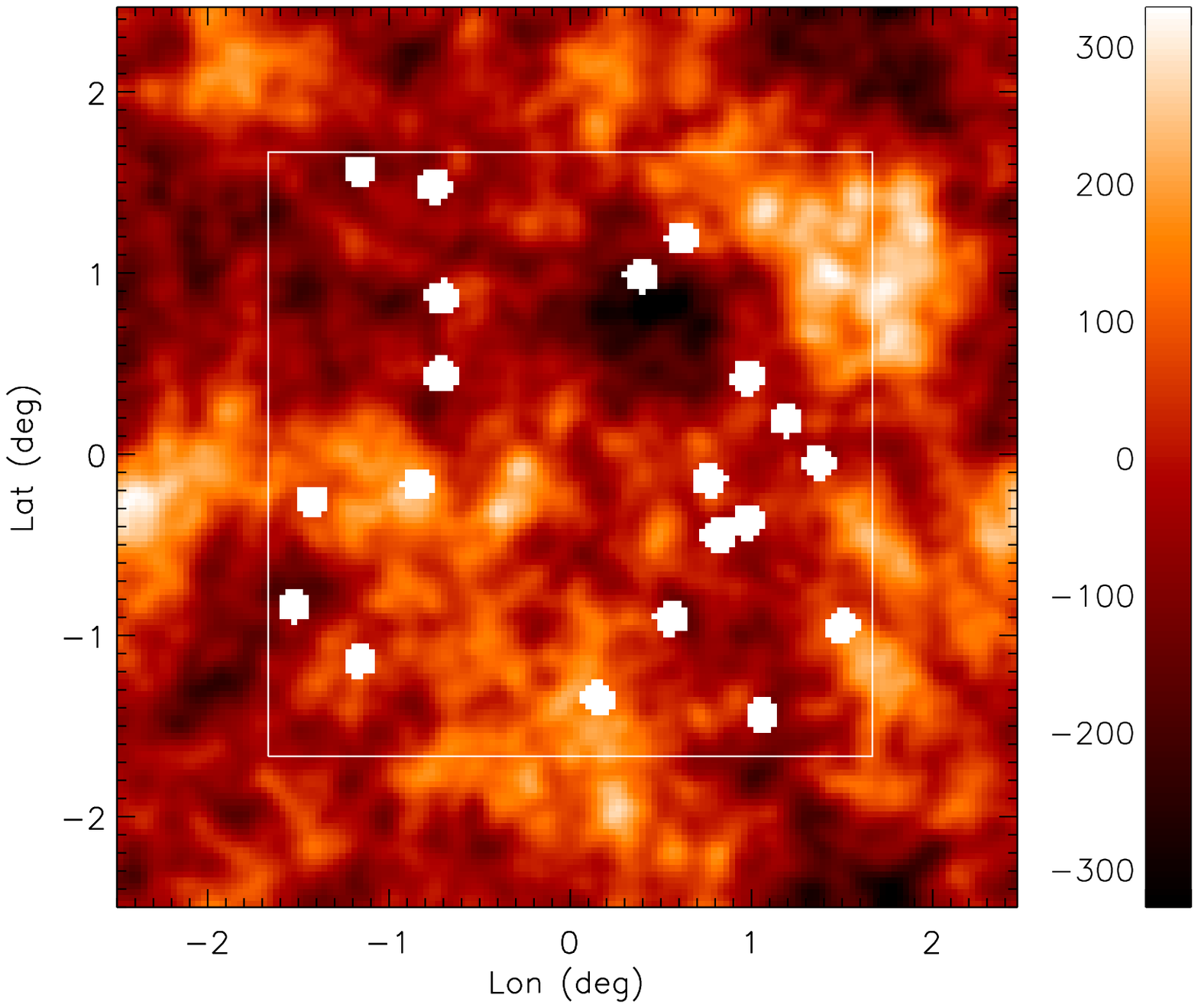}
\caption{Typical maps of dust with a $k^{-3}$ power spectrum (top) and CMB
  temperature (bottom). The square is the outline of the observed patch. It is
  extracted from a larger simulated map to ensure non-periodic boundary
  conditions. Masked data appear in white.}
\label{fig:maps}
\end{center}
\end{figure}

\section{Worked example \label{se:applis}}

\begin{figure}
\begin{center}
\includegraphics[clip, angle=0, scale = 0.3]{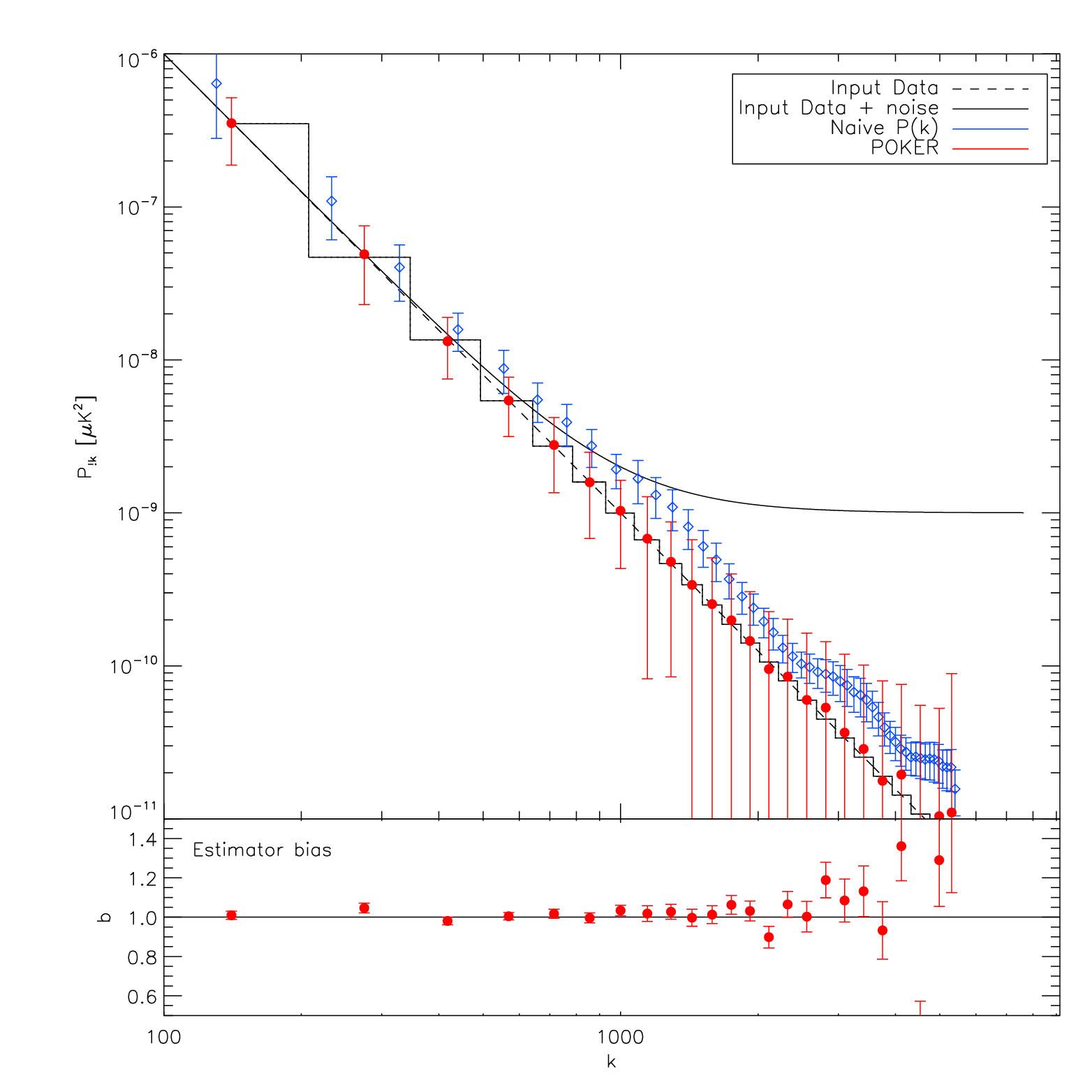}
\includegraphics[clip, angle=0, scale = 0.4]{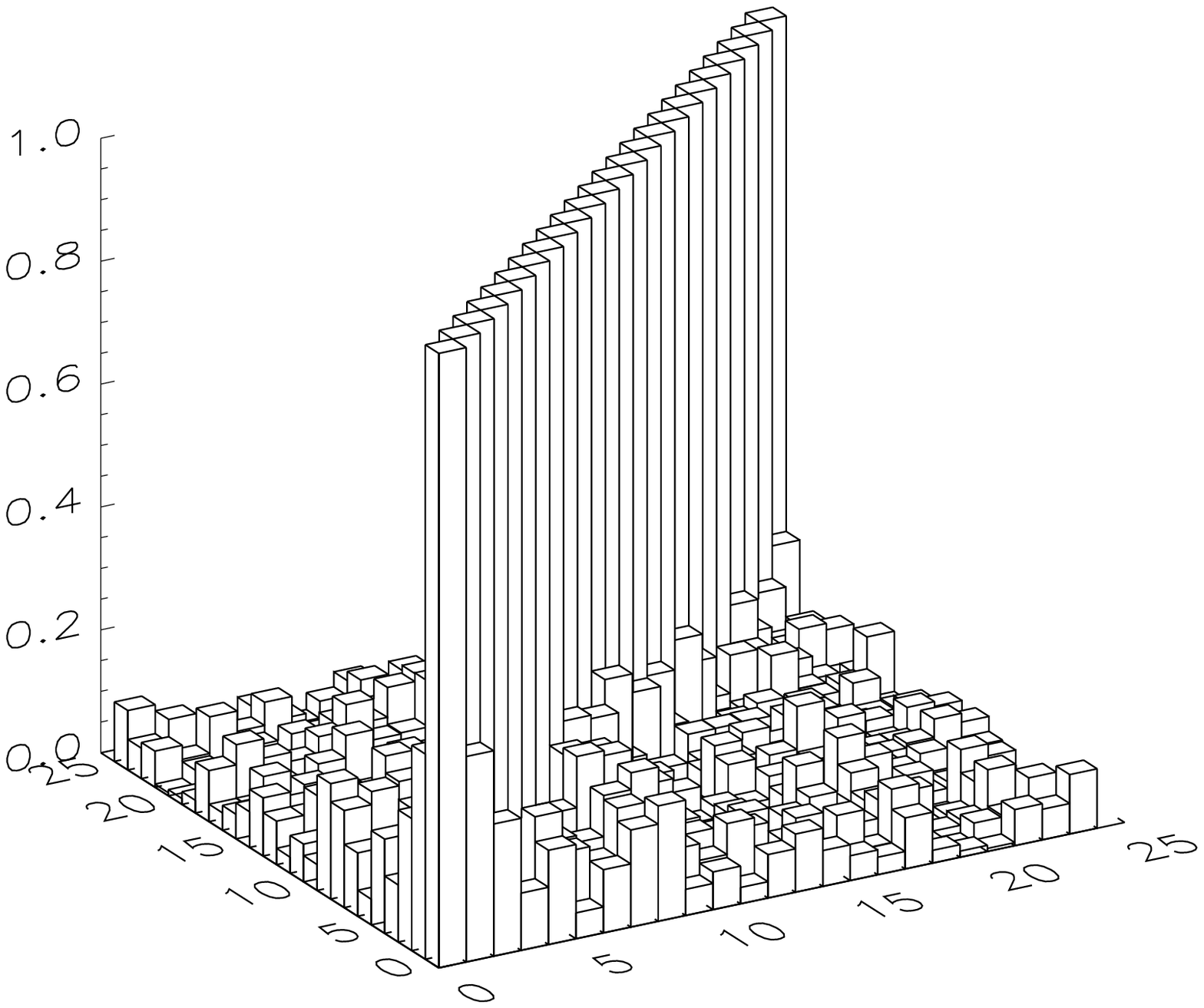}
\caption{\emph{Dust ($k^{-3}$)}. Comparison between the input theoretical power
  spectrum (black) and the average result of \poker~(red) applied to 500
  signal+noise simulations. The ``naive'' approach (blue, see text) is also
  shown for reference. Error bars in the top plots are those associated with the
  data (i.e. those of a single realization). The square line shows the binned
  theoretical power spectrum to which \poker's average result should be
  compared. The bottom plots shows the ratio of the reconstructed binned power
  spectrum to the input theoretical binned power spectrum (the bias) and the
  displayed error bar is that of the average of the Monte Carlo realization (in
  other words: the error bar of the top plot divided by $\sqrt{500}$). These
  plots altogether show that \poker~is unbiased. The mask used in this case is
  that of the top plot of Fig.~\ref{fig:masks}, with 1 where data are available,
  0 elsewhere.}
\label{fig:pk_dust_rawholes}
\end{center}
\end{figure}

\poker~was applied to real data to measure the cosmic infrared background
anisotropy in the Planck-HFI data \citep{planck_cib_2011}. The whole data
processing and how it is accounted for with \poker~is described in detail in
that paper. We do not replicate this analysis here but instead present
complementary examples on simulated data with steeper power spectra and a more
complex mask. It is in this context that mask aliasing effects are the
strongest.

\begin{figure*}
\begin{center}
\includegraphics[clip, angle=0, scale = 0.32]{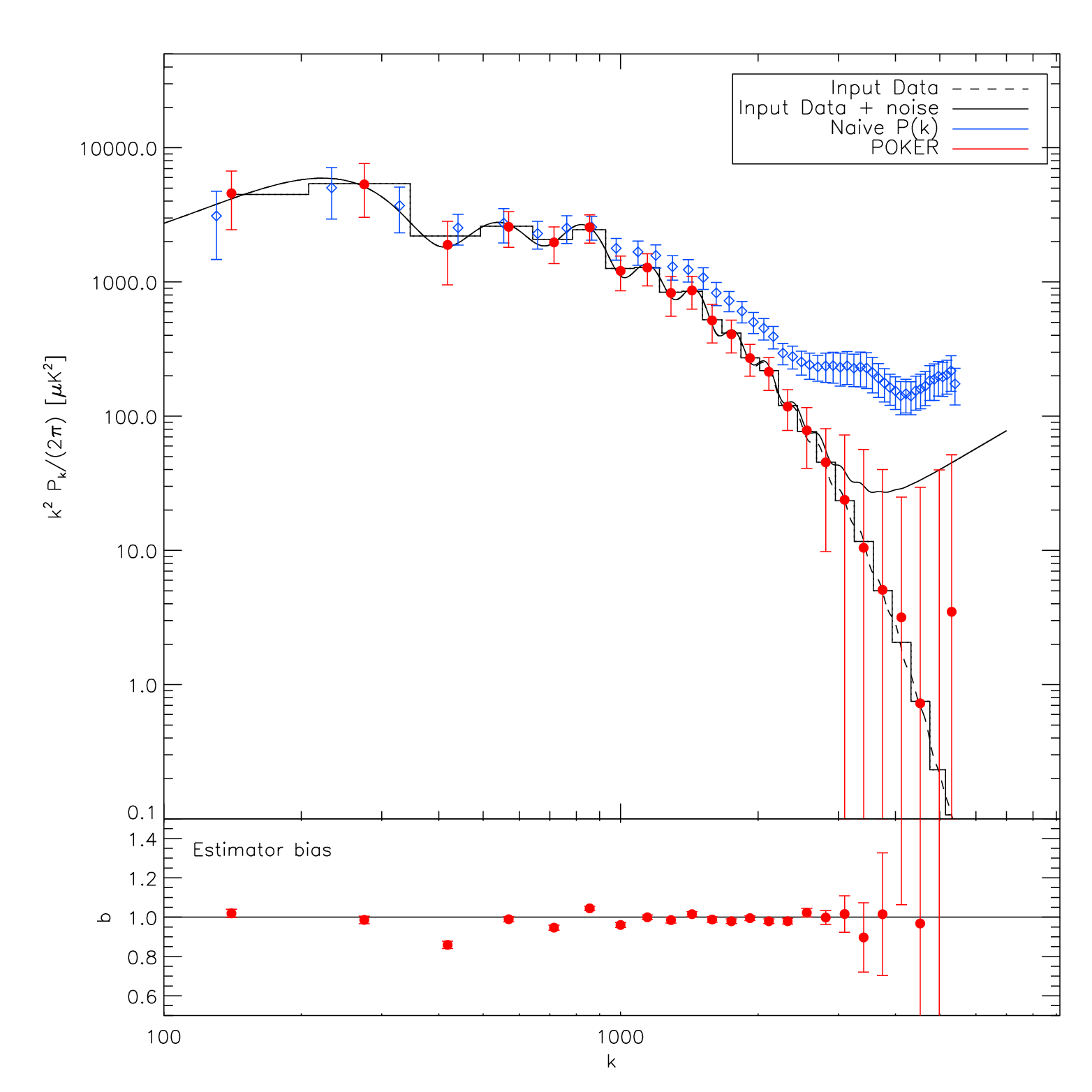}
\includegraphics[clip, angle=0, scale = 0.32]{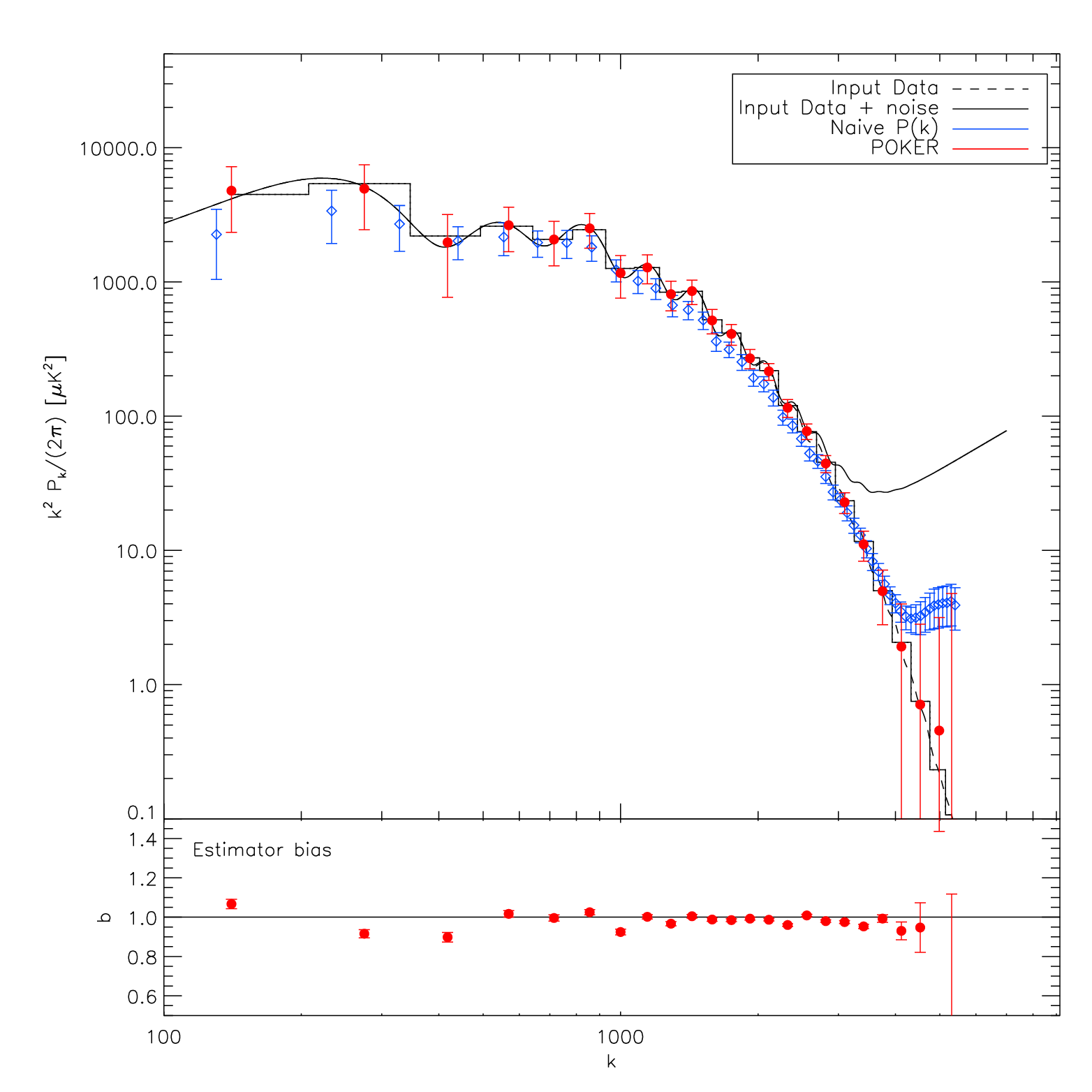}
\includegraphics[clip, angle=0, scale = 0.4]{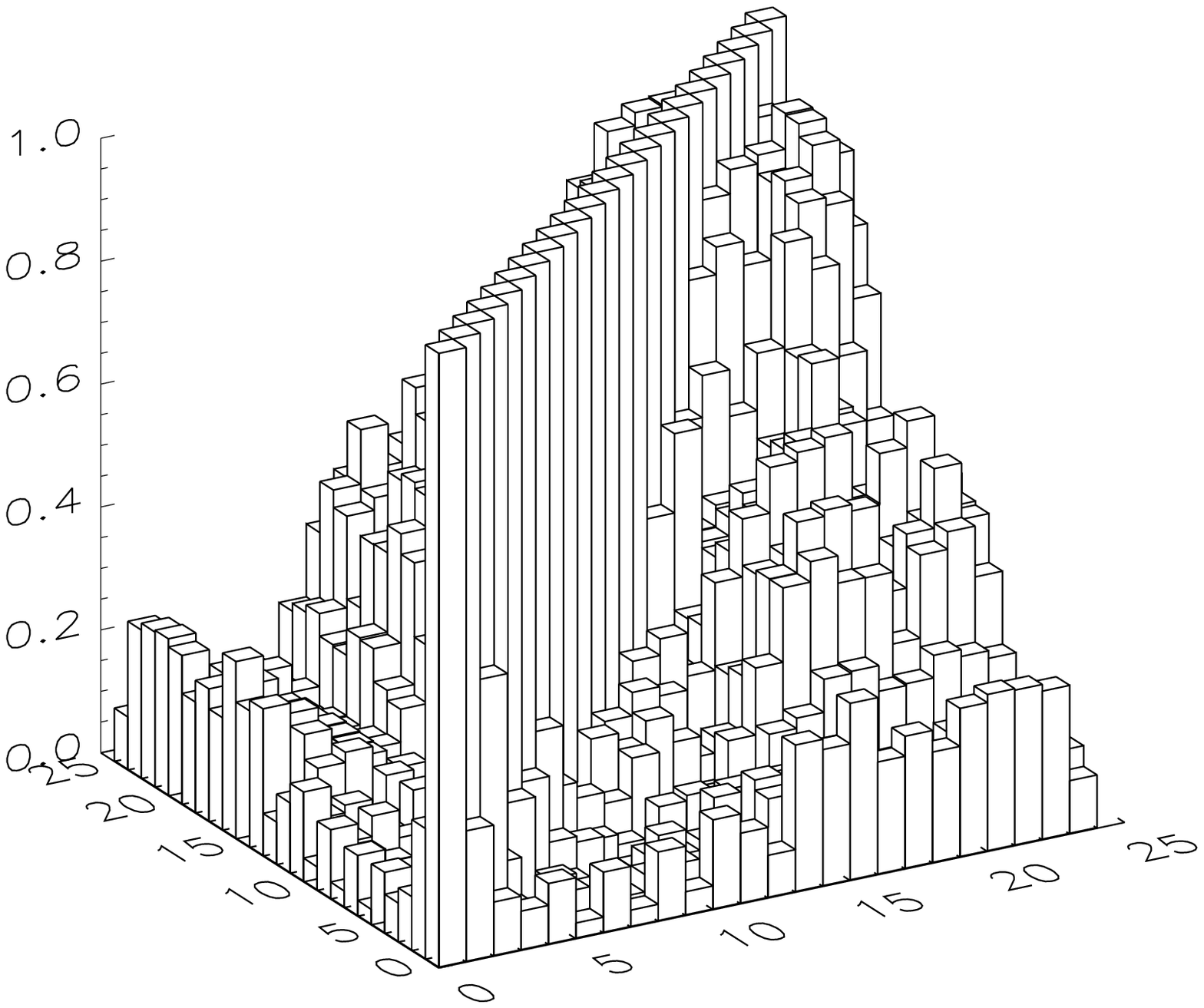}
\includegraphics[clip, angle=0, scale = 0.4]{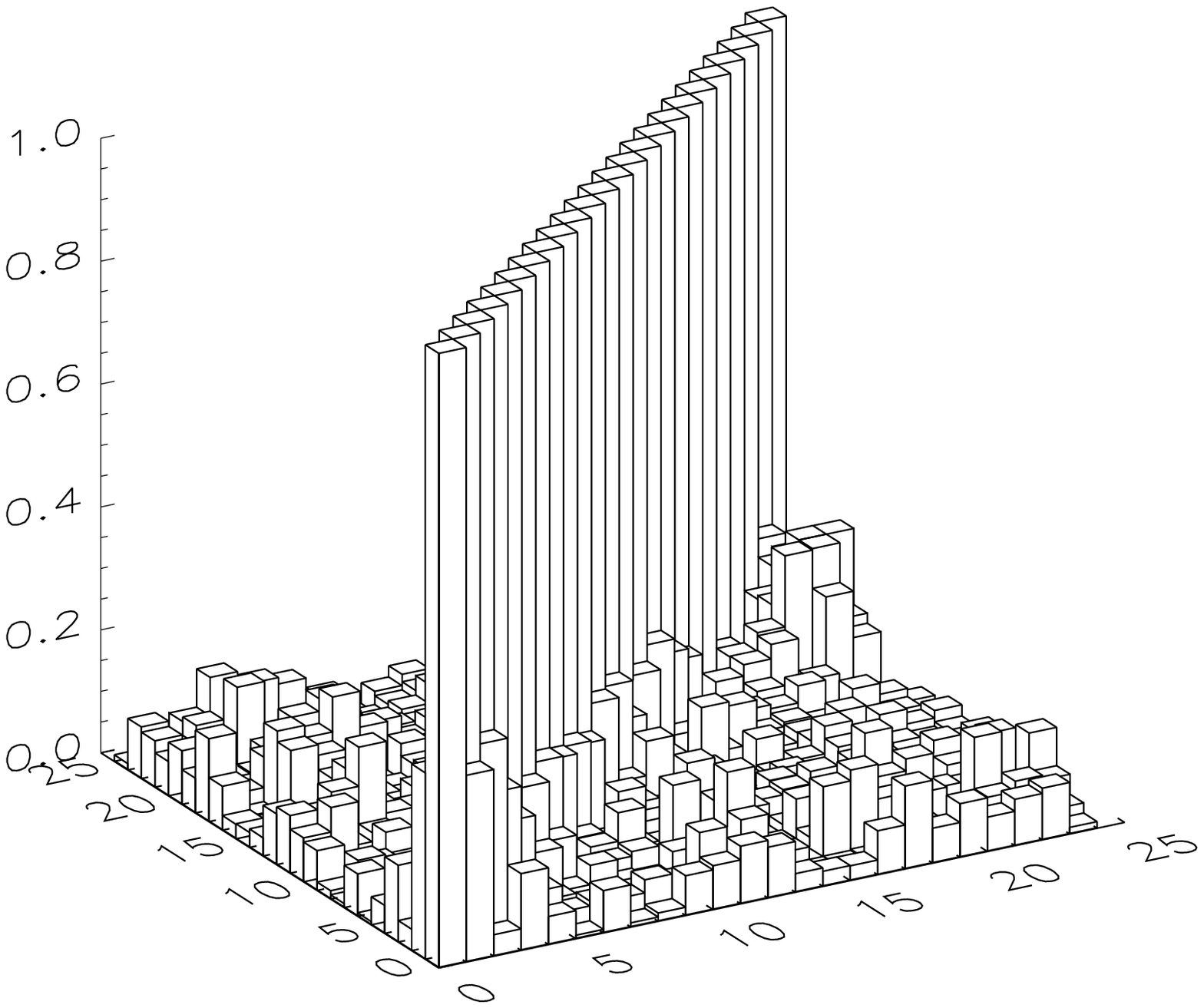}
\caption{\emph{Left}: same as Fig.~\ref{fig:pk_dust_rawholes} in the case of CMB. \poker's
  estimation is unbiased but mask aliasing induces strong correlations between
  bins at high $k$ where the power spectrum is very steep. \emph{Right}: This time, the mask with apodized
  boundaries (bottom of Fig.~\ref{fig:masks}) is used. High $k$ bin-to-bin correlations are significantly
  reduced. Apodization however reduces the effective observed fraction of the
  sky and therefore slightly increases error bars at low $k$ compared to the
  plots on the left. Note that although apodization also improves
  the ``naive'' estimate, it remains not compatible with the input spectrum for
  almost every bin and to more than the $1\sigma$ error of a single realization.}
\label{fig:pk_cmb_rawholes}
\end{center}
\end{figure*}

We assume that the observation patch is a square of 100 pixels of 2 arcmin
side. These parameters are chosen so that they sample a range of angular modes
over which the CMB temperature power spectrum exhibits peaks and a varying
slope. Note however that the map resolution can be arbitrary high because fast
Fourier transform algorithms work in dimensionless units. To force non-periodic
boundary conditions, we extract the patch from a map that is 50\% larger and the
simulation is performed on the latter. Finally, we draw random holes across the
observation patch to mimic point source masking. We consider two types of
signal. In the first case, we assume that the data are represented by a pure
power law spectrum $k^{-3}$ typical of Galactic dust emission. In the second
case, we assume that the data are CMB with a standard $\Lambda CDM$ power
spectrum. At these angular scales, the slope of the CMB power spectrum varies
from $\sim k^{-2}$ to even steeper than $k^{-6}$ and exhibits
oscillations. Fig.~\ref{fig:maps} shows an example of these simulated data. The
result of \poker~applied to each case is presented in
Figs.~\ref{fig:pk_dust_rawholes} and \ref{fig:pk_cmb_rawholes}. In the case of
dust, we choose a binning index $\beta=3$ as defined in Eq.~(\ref{eq:qmat}), and
in the case of CMB, we make no assumptions, \emph{i.e.} we choose $\beta=0$. We
also show what the direct Fourier transform of the observed patch without any
further correction would give to illustrate the magnitude of the effect
corrected by \poker. Note that this reference estimate labeled ``naive $P(k)$''
is not the \emph{pseudo}-power spectrum of the data in the sense of
Sect.~\ref{se:ps_def}. Indeed, it is not computed on the whole map from which
the observation patch is extracted and padded with zeros. The bottom plots of
Figs.~\ref{fig:pk_dust_rawholes} and \ref{fig:pk_cmb_rawholes} show the
bin-to-bin correlation matrix of each estimate. In the case of dust, the
correlations are small ($\sim 15\%$). This is not the case for the simulated
CMB, for which there is strong bin to bin correlation, although the power
spectrum remains unbiased. This correlation is due to large-scale aliasing
induced by the holes in the mask and show up so significantly because at high
$k$, the CMB spectrum is very steep. A way to improve on this is to apodize the
mask around the edges and the holes left by point-source masking
(Fig.~\ref{fig:masks}). In this work, we simply use a Gaussian kernel with a
FWHM of twice the map resolution to smooth the edges. The same analysis as
before is performed with this mask and results are presented in the right hand
side of Fig.~\ref{fig:pk_cmb_rawholes}. On this occasion, the bin-to-bin
correlation is significantly reduced, albeit the sampling variance is slighty
increased at low $k$ owing to the effective reduction in the observation area. A
more efficient way of performing this apodization is described in
\cite{xpure}. Finally, on larger angular scales, there is a slight bias in the
recovery of the CMB power spectrum. This does not however occur in the case of
dust, because for a pure power-law spectrum, Eq.~(\ref{eq:q}) is then an
equality. This is no longer the case for a CMB spectrum whose average slope
varies with $k$ and exhibits peaks. No binning could faithfully represent such a
spectrum. However, the remaining bias is negligible relative to the statistical
error bar in the data. If we force the simulated CMB to have a constant power
spectrum over frequency bins, the recovery is unbiased. There is no general
prescription regarding the definition of the binning and the apodization. They
must however be chosen with care because the bin-to-bin residual correlation may
lead to residual ringing (mask aliasing) in the data power spectrum (considered
as a single random realization), even if the estimator is, on average, unbiased.

\section{Conclusion}

We have developed a tool that provides an unbiased estimate of the angular power
spectrum of diffuse emission in the flat sky approximation limit, for arbitrary
high resolution and complex masks. \poker~corrects for mask aliasing effects,
even in the context of steep power spectra and provides a way to estimate
statistical error bars and bin-to-bin correlations. It complements tools
developed in the context of spherical sky and potentially full sky surveys
\citep[e.g.][]{master} but at the moment for lower angular
resolutions. \poker~also complements other methods in the flat sky
approximation such as \cite{das}.

\poker~can readily be generalized to polarization power-spectra estimation. To
date, experiments that have measured polarized diffuse emission
\citep{dasi,kogut,ponthieu2005,quad_pol2008,chiang2010,bierman2011} have been
closely related to CMB experiments and studied observation patches of a few to a
hundred percent of the sky and angular resolutions larger than a few
arcmin. Optimal tools have been developed to measure the polarization power
spectra in this context \citep[and references
  therein]{chon,smith2006,smith_zalda,xpure} and it is unlikely that \poker~will
bring something significantly new to the analysis of these observations. It is
however expected that smaller, deeper, and higher-resolution polarized surveys
will happen in the future, for which \poker~might be an interesting
approach. One of the main features that should then be addressed is the ability
of \poker~to correct for $E-B$ leakage. Although we postpone the detailed
studies of \poker's properties regarding polarized power spectra estimation to
future work, we provide the formalism in Appendices \ref{se:app_eb} and
\ref{se:app_te} for the sake of completeness. All the software used in this work
is publicly available\footnote{{\tt http://www.ias.u-psud.fr/poker}}.

\acknowledgements{We thank E.~Hivon, O.~Dor\'e, S.~Prunet, K.~Benabed
  and T.~Rodet for fruitful discussions. J.~Grain was supported by the ``Groupement
  d'Int\'er\^et Scientifique (GIS) Physique des 2 Infinis (P2I). This research
  used resources of the National Energy Research Scientific Computing Center,
  which is supported by the Office of Science of the U.S. Department of Energy
  under Contract No. DE-AC02-05CH11231.}

\bibliographystyle{aa}
\bibliography{references}

\begin{thebibliography}{23}
\expandafter\ifx\csname natexlab\endcsname\relax\def\natexlab#1{#1}\fi

\bibitem[{{Ade} {et~al.}(2008){Ade}, {Bock}, {Bowden}, {Brown}, {Cahill},
  {Carlstrom}, {Castro}, {Church}, {Culverhouse}, {Friedman}, {Ganga}, {Gear},
  {Hinderks}, {Kovac}, {Lange}, {Leitch}, {Melhuish}, {Murphy}, {Orlando},
  {Schwarz}, {O'Sullivan}, {Piccirillo}, {Pryke}, {Rajguru}, {Rusholme},
  {Taylor}, {Thompson}, {Wu}, \& {Zemcov}}]{quad_pol2008}
{Ade}, P., {Bock}, J., {Bowden}, M., {et~al.} 2008, \apj, 674, 22

\bibitem[{{Beno{\^i}t} {et~al.}(2003){Beno{\^i}t}, {Ade}, {Amblard}, {Ansari},
  {Aubourg}, {Bargot}, {Bartlett}, {Bernard}, {Bhatia}, {Blanchard}, {Bock},
  {Boscaleri}, {Bouchet}, {Bourrachot}, {Camus}, {Couchot}, {de Bernardis},
  {Delabrouille}, {D{\'e}sert}, {Dor{\'e}}, {Douspis}, {Dumoulin}, {Dupac},
  {Filliatre}, {Fosalba}, {Ganga}, {Gannaway}, {Gautier}, {Giard},
  {Giraud-H{\'e}raud}, {Gispert}, {Guglielmi}, {Hamilton}, {Hanany},
  {Henrot-Versill{\'e}}, {Kaplan}, {Lagache}, {Lamarre}, {Lange},
  {Mac{\'{\i}}as-P{\'e}rez}, {Madet}, {Maffei}, {Magneville}, {Marrone},
  {Masi}, {Mayet}, {Murphy}, {Naraghi}, {Nati}, {Patanchon}, {Perrin}, {Piat},
  {Ponthieu}, {Prunet}, {Puget}, {Renault}, {Rosset}, {Santos}, {Starobinsky},
  {Strukov}, {Sudiwala}, {Teyssier}, {Tristram}, {Tucker}, {Vanel}, {Vibert},
  {Wakui}, \& {Yvon}}]{archeops_t1}
{Beno{\^i}t}, A., {Ade}, P., {Amblard}, A., {et~al.} 2003, \aap, 399, L19

\bibitem[{{Bierman} {et~al.}(2011){Bierman}, {Matsumura}, {Dowell}, {Keating},
  {Ade}, {Barkats}, {Barron}, {Battle}, {Bock}, {Chiang}, {Culverhouse},
  {Duband}, {Hivon}, {Holzapfel}, {Hristov}, {Kaufman}, {Kovac}, {Kuo},
  {Lange}, {Leitch}, {Mason}, {Miller}, {Nguyen}, {Pryke}, {Richter}, {Rocha},
  {Sheehy}, {Takahashi}, \& {Yoon}}]{bierman2011}
{Bierman}, E.~M., {Matsumura}, T., {Dowell}, C.~D., {et~al.} 2011, ArXiv
  e-prints

\bibitem[{{Bond} {et~al.}(1998){Bond}, {Jaffe}, \& {Knox}}]{bjk}
{Bond}, J.~R., {Jaffe}, A.~H., \& {Knox}, L. 1998, \prd, 57, 2117

\bibitem[{{Brown} {et~al.}(2009){Brown}, {Ade}, {Bock}, {Bowden}, {Cahill},
  {Castro}, {Church}, {Culverhouse}, {Friedman}, {Ganga}, {Gear}, {Gupta},
  {Hinderks}, {Kovac}, {Lange}, {Leitch}, {Melhuish}, {Memari}, {Murphy},
  {Orlando}, {O'Sullivan}, {Piccirillo}, {Pryke}, {Rajguru}, {Rusholme},
  {Schwarz}, {Taylor}, {Thompson}, {Turner}, {Wu}, {Zemcov}, \& {The QUa D
  collaboration}}]{quad_t}
{Brown}, M.~L., {Ade}, P., {Bock}, J., {et~al.} 2009, \apj, 705, 978

\bibitem[{{Chiang} {et~al.}(2010){Chiang}, {Ade}, {Barkats}, {Battle},
  {Bierman}, {Bock}, {Dowell}, {Duband}, {Hivon}, {Holzapfel}, {Hristov},
  {Jones}, {Keating}, {Kovac}, {Kuo}, {Lange}, {Leitch}, {Mason}, {Matsumura},
  {Nguyen}, {Ponthieu}, {Pryke}, {Richter}, {Rocha}, {Sheehy}, {Takahashi},
  {Tolan}, \& {Yoon}}]{chiang2010}
{Chiang}, H.~C., {Ade}, P.~A.~R., {Barkats}, D., {et~al.} 2010, \apj, 711, 1123

\bibitem[{{Chon} {et~al.}(2004){Chon}, {Challinor}, {Prunet}, {Hivon}, \&
  {Szapudi}}]{chon}
{Chon}, G., {Challinor}, A., {Prunet}, S., {Hivon}, E., \& {Szapudi}, I. 2004,
  \mnras, 350, 914

\bibitem[{{Das} {et~al.}(2009){Das}, {Hajian}, \& {Spergel}}]{das}
{Das}, S., {Hajian}, A., \& {Spergel}, D.~N. 2009, \prd, 79, 083008

\bibitem[{{de Bernardis} {et~al.}(2000){de Bernardis}, {Ade}, {Bock}, {Bond},
  {Borrill}, {Boscaleri}, {Coble}, {Crill}, {De Gasperis}, {Farese},
  {Ferreira}, {Ganga}, {Giacometti}, {Hivon}, {Hristov}, {Iacoangeli}, {Jaffe},
  {Lange}, {Martinis}, {Masi}, {Mason}, {Mauskopf}, {Melchiorri}, {Miglio},
  {Montroy}, {Netterfield}, {Pascale}, {Piacentini}, {Pogosyan}, {Prunet},
  {Rao}, {Romeo}, {Ruhl}, {Scaramuzzi}, {Sforna}, \& {Vittorio}}]{boomerang}
{de Bernardis}, P., {Ade}, P.~A.~R., {Bock}, J.~J., {et~al.} 2000, \nat, 404,
  955

\bibitem[{{Gorski} {et~al.}(1999){Gorski}, {Wandelt}, {Hansen}, {Hivon}, \&
  {Banday}}]{healpix}
{Gorski}, K.~M., {Wandelt}, B.~D., {Hansen}, F.~K., {Hivon}, E., \& {Banday},
  A.~J. 1999, ArXiv Astrophysics e-prints

\bibitem[{{Grain} {et~al.}(2009){Grain}, {Tristram}, \& {Stompor}}]{xpure}
{Grain}, J., {Tristram}, M., \& {Stompor}, R. 2009, \prd, 79, 123515

\bibitem[{{Hinshaw} {et~al.}(2003){Hinshaw}, {Spergel}, {Verde}, {Hill},
  {Meyer}, {Barnes}, {Bennett}, {Halpern}, {Jarosik}, {Kogut}, {Komatsu},
  {Limon}, {Page}, {Tucker}, {Weiland}, {Wollack}, \& {Wright}}]{hinshaw}
{Hinshaw}, G., {Spergel}, D.~N., {Verde}, L., {et~al.} 2003, \apjs, 148, 135

\bibitem[{{Hivon} {et~al.}(2002){Hivon}, {G{\'o}rski}, {Netterfield}, {Crill},
  {Prunet}, \& {Hansen}}]{master}
{Hivon}, E., {G{\'o}rski}, K.~M., {Netterfield}, C.~B., {et~al.} 2002, \apj,
  567, 2

\bibitem[{{Kogut} {et~al.}(2003){Kogut}, {Spergel}, {Barnes}, {Bennett},
  {Halpern}, {Hinshaw}, {Jarosik}, {Limon}, {Meyer}, {Page}, {Tucker},
  {Wollack}, \& {Wright}}]{kogut}
{Kogut}, A., {Spergel}, D.~N., {Barnes}, C., {et~al.} 2003, \apjs, 148, 161

\bibitem[{{Kovac} {et~al.}(2002){Kovac}, {Leitch}, {Pryke}, {Carlstrom},
  {Halverson}, \& {Holzapfel}}]{dasi}
{Kovac}, J.~M., {Leitch}, E.~M., {Pryke}, C., {et~al.} 2002, \nat, 420, 772

\bibitem[{{Miville-Desch{\^e}nes} {et~al.}(2007){Miville-Desch{\^e}nes},
  {Lagache}, {Boulanger}, \& {Puget}}]{mamd_07}
{Miville-Desch{\^e}nes}, M.-A., {Lagache}, G., {Boulanger}, F., \& {Puget},
  J.-L. 2007, \aap, 469, 595

\bibitem[{{Planck Collaboration} {et~al.}(2011){Planck Collaboration}, {Ade},
  {Aghanim}, {Arnaud}, {Ashdown}, {Aumont}, {Baccigalupi}, {Balbi}, {Banday},
  {Barreiro}, {Bartlett}, {Battaner}, {Benabed}, {Benoit}, {Bernard},
  {Bersanelli}, {Bhatia}, {Blagrave}, {Bock}, {Bonaldi}, {Bonavera}, {Bond},
  {Borrill}, {Bouchet}, {Bucher}, {Burigana}, {Cabella}, {Cardoso}, {Catalano},
  {Cayon}, {Challinor}, {Chamballu}, {Chiang}, {Chiang}, {Christensen},
  {Clements}, {Colombi}, {Couchot}, {Coulais}, {Crill}, {Cuttaia}, {Danese},
  {Davies}, {Davis}, {de Bernardis}, {de Gasperis}, {de Rosa}, {de Zotti},
  {Delabrouille}, {Delouis}, {Desert}, {Dole}, {Donzelli}, {Dore}, {Dorl},
  {Douspis}, {Dupac}, {Efstathiou}, {Ensslin}, {Eriksen}, {Finelli}, {Forni},
  {Fosalba}, {Frailis}, {Franceschi}, {Galeotta}, {Ganga}, {Giard}, {Giardino},
  {Giraud-Heraud}, {Gonzalez-Nuevo}, {Gorski}, {Grain}, {Gratton}, {Gregorio},
  {Gruppuso}, {Hansen}, {Harrison}, {Helou}, {Henrot-Versille}, {Herranz},
  {Hildebrandt}, {Hivon}, {Hobson}, {Holmes}, {Hovest}, {Hoyland},
  {Huffenberger}, {Jaffe}, {Jones}, {Juvela}, {Keihanen}, {Keskitalo},
  {Kisner}, {Kneissl}, {Knox}, {Kurki-Suonio}, {Lagache}, {Lamarre}, {Lasenby},
  {Laureijs}, {Lawrence}, {Leach}, {Leonardi}, {Leroy}, {Lilje},
  {Linden-Vornle}, {Lockman}, {Lopez-Caniego}, {Lubin}, {Macias-Perez},
  {MacTavish}, {Maffei}, {Maino}, {Mandolesi}, {Mann}, {Maris}, {Martin},
  {Martinez-Gonzalez}, {Masi}, {Matarrese}, {Matthai}, {Mazzotta},
  {Melchiorri}, {Mendes}, {Mennella}, {Mitra}, {Miville-Deschenes}, {Moneti},
  {Montier}, {Morgante}, {Mortlock}, {Munshi}, {Murphy}, {Naselsky}, {Natoli},
  {Netterfield}, {Norgaard-Nielsen}, {Novikov}, {Novikov}, {O'Dwyer}, {Oliver},
  {Osborne}, {Pajot}, {Pasian}, {Patanchon}, {Perdereau}, {Perotto},
  {Perrotta}, {Piacentini}, {Piat}, {Pinheiro Goncalves}, {Plaszczynski},
  {Pointecouteau}, {Polenta}, {Ponthieu}, {Poutanen}, {Prezeau}, {Prunet},
  {Puget}, {Rachen}, {Reach}, {Reinecke}, {Remazeilles}, {Renault},
  {Ricciardi}, {Riller}, {Ristorcelli}, {Rocha}, {Rosset}, {Rowan-Robinson},
  {Rubino-Martin}, {Rusholme}, {Sandri}, {Santos}, {Savini}, {Scott},
  {Seiffert}, {Shellard}, {Smoot}, {Starck}, {Stivoli}, {Stolyarov}, {Stompor},
  {Sudiwala}, {Sunyaev}, {Sygnet}, {Tauber}, {Terenzi}, {Toffolatti}, {Tomasi},
  {Torre}, {Tristram}, {Tuovinen}, {Umana}, {Valenziano}, {Vielva}, {Villa},
  {Vittorio}, {Wade}, {Wandelt}, {White}, {Yvon}, {Zacchei}, \&
  {Zonca}}]{planck_cib_2011}
{Planck Collaboration}, {Ade}, P.~A.~R., {Aghanim}, N., {et~al.} 2011, ArXiv
  e-prints 1101.2028

\bibitem[{{Ponthieu} {et~al.}(2005){Ponthieu}, {Mac{\'{\i}}as-P{\'e}rez},
  {Tristram}, {Ade}, {Amblard}, {Ansari}, {Aumont}, {Aubourg}, {Beno{\^i}t},
  {Bernard}, {Blanchard}, {Bock}, {Bouchet}, {Bourrachot}, {Camus}, {Cardoso},
  {Couchot}, {de Bernardis}, {Delabrouille}, {D{\'e}sert}, {Douspis},
  {Dumoulin}, {Filliatre}, {Fosalba}, {Giard}, {Giraud-H{\'e}raud}, {Gispert},
  {Grain}, {Guglielmi}, {Hamilton}, {Hanany}, {Henrot-Versill{\'e}}, {Kaplan},
  {Lagache}, {Lange}, {Madet}, {Maffei}, {Masi}, {Mayet}, {Nati}, {Patanchon},
  {Perdereau}, {Plaszczynski}, {Piat}, {Prunet}, {Puget}, {Renault}, {Rosset},
  {Santos}, {Vibert}, \& {Yvon}}]{ponthieu2005}
{Ponthieu}, N., {Mac{\'{\i}}as-P{\'e}rez}, J.~F., {Tristram}, M., {et~al.}
  2005, \aap, 444, 327

\bibitem[{{Pryke} {et~al.}(2009){Pryke}, {Ade}, {Bock}, {Bowden}, {Brown},
  {Cahill}, {Castro}, {Church}, {Culverhouse}, {Friedman}, {Ganga}, {Gear},
  {Gupta}, {Hinderks}, {Kovac}, {Lange}, {Leitch}, {Melhuish}, {Memari},
  {Murphy}, {Orlando}, {Schwarz}, {O'Sullivan}, {Piccirillo}, {Rajguru},
  {Rusholme}, {Taylor}, {Thompson}, {Turner}, {Wu}, \& {Zemcov}}]{pryke}
{Pryke}, C., {Ade}, P., {Bock}, J., {et~al.} 2009, \apj, 692, 1247

\bibitem[{{Reichardt} {et~al.}(2009){Reichardt}, {Ade}, {Bock}, {Bond},
  {Brevik}, {Contaldi}, {Daub}, {Dempsey}, {Goldstein}, {Holzapfel}, {Kuo},
  {Lange}, {Lueker}, {Newcomb}, {Peterson}, {Ruhl}, {Runyan}, \&
  {Staniszewski}}]{acbar}
{Reichardt}, C.~L., {Ade}, P.~A.~R., {Bock}, J.~J., {et~al.} 2009, \apj, 694,
  1200

\bibitem[{{Smith}(2006)}]{smith2006}
{Smith}, K.~M. 2006, \nar, 50, 1025

\bibitem[{{Smith} \& {Zaldarriaga}(2007)}]{smith_zalda}
{Smith}, K.~M. \& {Zaldarriaga}, M. 2007, \prd, 76, 043001

\bibitem[{{Tristram} {et~al.}(2005){Tristram}, {Patanchon},
  {Mac{\'{\i}}as-P{\'e}rez}, {Ade}, {Amblard}, {Ansari}, {Aubourg},
  {Beno{\^i}t}, {Bernard}, {Blanchard}, {Bock}, {Bouchet}, {Bourrachot},
  {Camus}, {Cardoso}, {Couchot}, {de Bernardis}, {Delabrouille}, {D{\'e}sert},
  {Douspis}, {Dumoulin}, {Filliatre}, {Fosalba}, {Giard}, {Giraud-H{\'e}raud},
  {Gispert}, {Guglielmi}, {Hamilton}, {Hanany}, {Henrot-Versill{\'e}},
  {Kaplan}, {Lagache}, {Lamarre}, {Lange}, {Madet}, {Maffei}, {Magneville},
  {Masi}, {Mayet}, {Nati}, {Perdereau}, {Plaszczynski}, {Piat}, {Ponthieu},
  {Prunet}, {Renault}, {Rosset}, {Santos}, {Vibert}, \& {Yvon}}]{archeops_t2}
{Tristram}, M., {Patanchon}, G., {Mac{\'{\i}}as-P{\'e}rez}, J.~F., {et~al.}
  2005, \aap, 436, 785

\end{thebibliography}

\newpage
\onecolumn

\begin{appendix}

\section{Mask convolution kernel for temperature \label{se:fsl_dft}}

If not specified, sums run from 0 to $N_x-1$ and 0 to $N_y-1$. The
\emph{pseudo}-Fourier coefficients of the weighted data are given by

\begin{eqnarray}
\hat{T}_{mn} & = &
\frac{1}{N_xN_y}\sum_{\mu,\nu}T_{\mu\nu}W_{\mu\nu}e^{-2i\pi(\mu m/N_x+\nu n/N_y)}, \label{eq:raw_ft} \\
& = & \frac{1}{N_xN_y}\sum_{\mu,\nu}\sum_{m_1,n_1} T_{m_1n_1}e^{2i\pi(\mu m_1/N_x+\nu n_1/N_y)}
\sum_{m_2,n_2}W_{m_2n_2}e^{2i\pi(\mu m_2/N_x+\nu n_2/N_y)}e^{-2i\pi(\mu m/N_x+\nu n/N_y)}, \nonumber \\
& = & \frac{1}{N_xN_y}\sum_{m_1,n_1,m_2,n_2}\sum_{\mu,\nu}T_{m_1n_1}W_{m_2n_2}e^{2i\pi[\mu(m_1N_x+m_2-m)/N_x+\nu(n_1+n_2-n)/N_y]}.
\end{eqnarray}

\noindent $m_1+m_2$ belongs to $[0,2N_x-2]$, so
\begin{equation}
\sum_{\mu=0}^{N_x-1}e^{2i\pi \mu(m_1+m_2-m)/N_x} = N_x\delta_{m_2}^{m-m_1} + N_x\delta_{m_2}^{N_x+m-m_1}.
\end{equation}

\noindent Similar relations hold for indices $n$, hence

\begin{equation}
\hat{T}_{mn} = \sum_{m_1,n_1,m_2,n_2}T_{m_1n_1}W_{m_2n_2}\left(\delta_{m_2}^{m-m_1}
+ \delta_{m_2}^{N_x+m-m_1}\right)\left(\delta_{n_2}^{n-n_1}+
\delta_{n_2}^{N_y+n-n_1}\right).
\end{equation}

Equation (\ref{eq:dft}) specifies that Fourier coefficients are only defined for
$m,n \in [0,N_x-1] \times [0,N_y-1]$, hence

\begin{equation}
\hat{T}_{mn} =  \sum_{m_1,n_1}T_{m_1,n_1}K_{m,m_1}^{n,n_1}\;,
\label{eq:tw}
\end{equation}

\noindent where

\begin{equation}                                                                                                                                
K_{m,m_1}^{n,n_1} =
\left\{ \begin{array}{ll}
W_{m-m_1,\,n-n_1}        &\;{\rm if}\; m_1 \leq m \;{\rm and}\; n_1\leq n \\
W_{m-m_1,\,N_y+n-n_1}    &\;{\rm if}\; m_1 \leq m \;{\rm and}\; n_1> n \\
W_{Nx+m-m_1,\,n-n_1}     &\;{\rm if}\; m_1 > m    \;{\rm and}\; n_1 \leq n \\
W_{Nx+m-m_1,\,N_y+n-n_1} &\;{\rm if}\; m_1 > m    \;{\rm and}\; n_1 > n
\end{array}\right. .
\end{equation}

\section{Mask convolution kernel for polarization only \label{se:app_eb}}

Polarization maps are represented in direct space by Stokes parameters $Q$ and
$U$, in angular space by $E$ and $B$. These parameters are related by

\begin{eqnarray}
	Q_{\mu\nu}&=&\displaystyle\sum_{m_1,n_1}\left[\cos(2\phi_{m_1n_1}^{\mu\nu})E_{m_1n_1}-\sin(2\phi_{m_1n_1}^{\mu\nu})B_{m_1n_1}\right]e^{2i\pi(\mu
          m_1/N_x+\nu n_1/N_y)}\\
	U_{\mu\nu}&=&\displaystyle\sum_{m_1,n_1}\left[\sin(2\phi_{m_1n_1}^{\mu\nu})E_{m_1n_1}+\cos(2\phi_{m_1n_1}^{\mu\nu})B_{m_1n_1}\right]e^{2i\pi(\mu m_1/N_x+\nu n_1/N_y)},
\end{eqnarray}

\noindent The polarization \emph{pseudo}-Fourier coefficients are given by
\begin{eqnarray}
	\hat{E}_{mn}&=&\frac{1}{N_xN_y}\displaystyle\sum_{\mu,\nu}\left[\cos(2\phi_{mn}^{\mu\nu})Q_{\mu\nu}+\sin(2\phi_{mn}^{\mu\nu})U_{\mu\nu}\right]W_{\mu\nu}e^{-2i\pi(\mu m/N_x+\nu n/N_y)}
        , \\
&=&\frac{1}{N_xN_y}\displaystyle\sum_{m_1,n_1}\sum_{m_2,n_2}\sum_{\mu,\nu}W^{n_2}_{m_2}\left[\cos{(2\phi_{mn}^{\mu\nu}-2\phi_{m_1n_1}^{\mu\nu})}E_{m_1n_1}+\sin{(2\phi_{mn}^{\mu\nu}-2\phi_{m_1n_1}^{\mu\nu})}B_{m_1n_1}\right] \\
	&&\times e^{2i\pi(\mu m_1/N_x+\nu n_1/N_y)}e^{2i\pi(\mu m_2/N_x+\nu n_2/N_y)}e^{-2i\pi(\mu m/N_x+\nu n/N_y)}, \nonumber \\
	\hat{B}_{mn}&=&\frac{1}{N_xN_y}\displaystyle\sum_{\mu,\nu}\left[-\sin(2\phi_{mn}^{\mu\nu})Q_{\mu\nu}+\cos(2\phi_{mn}^{\mu\nu})U_{\mu\nu}\right]W_{\mu\nu}e^{-2i\pi(\mu m/N_x+\nu n/N_y)}
        ,\\
&=&\frac{1}{N_xN_y}\displaystyle\sum_{m_1,n_1}\sum_{m_2,n_2}\sum_{\mu,\nu}W^{n_2}_{m_2}\left[-\sin{(2\phi_{mn}^{\mu\nu}-2\phi_{m_1n_1}^{\mu\nu})}E_{m_1n_1}+\cos{(2\phi_{mn}^{\mu\nu}-2\phi_{m_1n_1}^{\mu\nu})}B_{m_1n_1}\right] \\
	&&\times e^{2i\pi(\mu m_1/N_x+\nu n_1/N_y)}e^{2i\pi(\mu m_2/N_x+\nu n_2/N_y)}e^{-2i\pi(\mu m/N_x+\nu n/N_y)}. \nonumber
\end{eqnarray}

\noindent We now have to compute the two summations
\begin{eqnarray}
	I_1&=&\displaystyle\sum_{\mu,\nu}\cos{(2\phi_{mn}^{\mu\nu}-2\phi_{m_1n_1}^{\mu\nu})}e^{2i\pi(\mu m_1/N_x+\nu n_1/N_y)}e^{2i\pi(\mu m_2/N_x+\nu n_2/N_y)}e^{-2i\pi(\mu m/N_x+\nu n/N_y)}, \\
	I_2&=&\displaystyle\sum_{\mu,\nu}\sin{(2\phi_{mn}^{\mu\nu}-2\phi_{m_1n_1}^{\mu\nu})}e^{2i\pi(\mu m_1/N_x+\nu n_1/N_y)}e^{2i\pi(\mu m_2/N_x+\nu n_2/N_y)}e^{-2i\pi(\mu m/N_x+\nu n/N_y)}.
\end{eqnarray}

\noindent Because $\phi_{mn}^{\mu\nu}$ is the angle between $\vec{k}_{mn}$ and
$\vec{r}_{\mu\nu}$ and $\phi_{m_1n_1}^{\mu\nu}$ is the angle between
$\vec{k}_{m_1n_1}$ and $\vec{r}_{\mu\nu}$, we have
$\phi_{mn}^{\mu\nu}-\phi_{m_1n_1}^{\mu\nu}=\phi_{m_1n_1}^{mn}$ with
$$
\cos(\phi_{mn}^{m_1n_1})=\vec{k}_{m_1n_1}\cdot\vec{k}_{mn}.
$$ As a consequence, the sine and cosine do not depend on pixel indices, $\mu$ and
$\nu$, so we can use the orthogonality relation used for temperature,
i.e. $\forall$ $m_1+m_2$ belongs to $[0,2N-2]$ :

$$
\sum_{\mu=0}^{N-1}e^{2i\pi \mu(m_1+m_2-m)/N_x} = N_x\delta_{m_2}^{m-m_1} + N_x\delta_{m_2}^{N+m-m_1},
$$

\noindent to finally get 

\begin{eqnarray}
	\hat{E}_{mn}&=&\displaystyle\sum_{m_1,n_1}K_{m,m_1}^{n,n_1}\left[\cos{(2\phi_{mn}^{m_1n_1})}E_{m_1n_1}+\sin{(2\phi_{mn}^{m_1n_1})}B_{m_1n_1}\right], \\
	\hat{B}_{mn}&=&\displaystyle\sum_{m_1,n_1}K_{m,m_1}^{n,n_1}\left[-\sin{(2\phi_{mn}^{m_1n_1})}E_{m_1n_1}+\cos{(2\phi_{mn}^{m_1n_1})}B_{m_1n_1}\right].
\end{eqnarray}

\noindent From these last set of results, we can compute the \emph{pseudo}-power spectra, keeping in mind
that

$$\left<E^{n}_{m}E_{m'n'}^*\right>=C^{EE}_{mn}\delta_{m,m'}\delta_{n,n'},$$
$$\left<B^{n}_{m}B_{m'n'}^*\right>=C^{BB}_{mn}\delta_{m,m'}\delta_{n,n'}$$

\noindent and

$$\left<E^{n}_{m}B_{m'n'}^*\right>=\left<B^{n}_{m}E_{m'n'}^*\right>=C^{EB}_{mn}\delta_{m,m'}\delta_{n,n'}.$$

\noindent We define the estimated \emph{pseudo}-spectra as
$$\hat{P}^{EE}_{mn}=\left|\hat{E}_{mn}\right|^2,$$
$$\hat{C^{BB}}_{mn}=\left|\hat{B}_{mn}\right|^2$$

\noindent and
$$\hat{P}^{EB}_{mn}=\displaystyle\frac{1}{2}\left[\hat{E}_{mn}\hat{B}_{mn}^*+\hat{B}_{mn}\hat{E}_{mn}^*\right]=\mathrm{Re}\left[\hat{E}_{mn}\hat{B}_{mn}^*\right]=\mathrm{Re}\left[\hat{B}_{mn}\hat{E}_{mn}^*\right].$$

\noindent By using those definitions, we can easily show that

\begin{eqnarray}
	\hat{P}^{EE}_{mn}&=&\displaystyle\sum_{m_1,n_1}|K_{m,m_1}^{n,n_1}|^2
	\left[\cos^2{(2\phi_{mn}^{m_1n_1})}C^{EE}_{m_1n_1}+\sin^2{(2\phi_{mn}^{m_1n_1})}C^{BB}_{m_1n_1}+\sin{(4\phi_{mn}^{m_1n_1})}C^{EB}_{m_1n_1}\right], \nonumber \\
	\hat{P}^{BB}_{mn}&=&\displaystyle\sum_{m_1,n_1}|K_{m,m_1}^{n,n_1}|^2
\left[\sin^2{(2\phi_{mn}^{m_1n_1})}C^{EE}_{m_1n_1}+\cos^2{(2\phi_{mn}^{m_1n_1})}C^{BB}_{m_1n_1}-\sin{(4\phi_{mn}^{m_1n_1})}C^{EB}_{m_1n_1}\right], \nonumber \\
	\hat{P}^{EB}_{mn}&=&\displaystyle\sum_{m_1,n_1}|K_{m,m_1}^{n,n_1}|^2
\left[-\frac{1}{2}\sin{(4\phi_{mn}^{m_1n_1})}C^{EE}_{m_1n_1}+\frac{1}{2}\sin{(4\phi_{mn}^{m_1n_1})}C^{BB}_{m_1n_1}+\left(\cos^2{(2\phi_{mn}^{m_1n_1})}-\sin^2{(2\phi_{mn}^{m_1n_1})}\right)C^{EB}_{m_1n_1}\right]. \nonumber
\end{eqnarray}

\noindent In a matrix formulation, this reads
\begin{eqnarray}
	\left(\begin{array}{c}
		\hat{P}^{EE}_{mn} \\
		\hat{P}^{BB}_{mn} \\
		\hat{P}^{EB}_{mn}
	\end{array}\right)&=&\displaystyle\sum_{m_1,n_1}\left(\begin{array}{ccc}
		M^{EE,EE}_{mn,m_1n_1} & M^{EE,BB}_{mn,m_1n_1} & M^{EE,EB}_{mn,m_1n_1} \\
		M^{BB,EE}_{mn,m_1n_1} & M^{BB,BB}_{mn,m_1n_1} & M^{BB,EB}_{mn,m_1n_1} \\
		M^{EB,EE}_{mn,m_1n_1} & M^{EB,BB}_{mn,m_1n_1} & M^{EB,EB}_{mn,m_1n_1}
	\end{array}\right)\left(\begin{array}{c}
		{P}^{{EE}}_{m_1n_1} \\
		{P}^{{BB}}_{m_1n_1} \\
		{P}^{{EB}}_{m_1n_1}
	\end{array}\right) \\
	&=&\displaystyle\sum_{m_1,n_1}\left(\begin{array}{ccc}
		M^{diag}_{mn,m_1n_1} & M^{off}_{mn,m_1n_1} & M^{cross}_{mn,m_1n_1} \\
		M^{off}_{mn,m_1n_1} & M^{off}_{mn,m_1n_1} & -M^{cross}_{mn,m_1n_1} \\
		-\frac{1}{2}M^{cross}_{mn,m_1n_1} & \frac{1}{2}M^{cross}_{mn,m_1n_1} & M^{diag}_{mn,m_1n_1}-M^{off}_{mn,m_1n_1}
	\end{array}\right)\left(\begin{array}{c}
		{P}^{{EE}}_{m_1n_1} \\
		{P}^{{BB}}_{m_1n_1} \\
		{P}^{{EB}}_{m_1n_1}
	\end{array}\right),
\end{eqnarray}

\noindent where

\begin{eqnarray}
	M^{diag}_{mn,m_1n_1}&=&\cos^2{(2\phi_{mn}^{m_1n_1})}|K_{m,m_1}^{n,n_1}|^2, \\
	M^{off}_{mn,m_1n_1}&=&\sin^2{(2\phi_{mn}^{m_1n_1})}|K_{m,m_1}^{n,n_1}|^2, \\
	M^{cross}_{mn,m_1n_1}&=&\sin{(4\phi_{mn}^{m_1n_1})}|K_{m,m_1}^{n,n_1}|^2.
\end{eqnarray}

\noindent When the above mixing matrices are averaged over the two azimuthal (or polar for
flat sky) angles, it can be shown that $\int\int M^{cross}_{mn,m_1n_1}d\theta
d\theta_1=0$. However, before such an averaging, it is not {\it a priori} zero.

\section{Mask convolution kernel for temperature polarization
  cross-correlation \label{se:app_te}}
For the cross-correlation of the temperature with CMB maps, we remind first that

$$\left<T_{mn}E_{m'n'}^*\right>=\left<E_{mn}T_{m'n'}^*\right>=C^{TE}_{mn}\delta_{m,m'}\delta_{n,n'}$$

\noindent and

$$\left<T_{mn}B_{m'n'}^*\right>=\left<B_{mn}T_{m'n'}^*\right>=C^{TB}_{mn}\delta_{m,m'}\delta_{n,n'}.$$

\noindent The estimated cross-\emph{pseudo}-spectrum are defined as

$$\hat{P}^{TE}_{mn}=\displaystyle\frac{1}{2}\left[\hat{T}_{mn}\hat{E}_{mn}^*+\hat{E}_{mn}\hat{T}_{mn}^*\right]=\mathrm{Re}\left[\hat{T}_{mn}\hat{E}_{mn}^*\right]=\mathrm{Re}\left[\hat{E}_{mn}\hat{T}_{mn}^*\right]$$

\noindent and

$$\hat{P}^{TB}_{mn}=\displaystyle\frac{1}{2}\left[\hat{T}_{mn}\hat{B}_{mn}^*+\hat{B}_{mn}\hat{T}_{mn}^*\right]=\mathrm{Re}\left[\hat{T}_{mn}\hat{B}_{mn}^*\right]=\mathrm{Re}\left[\hat{B}_{mn}\hat{T}_{mn}^*\right].$$

\noindent From this and from the above defined \emph{pseudo}-Fourier coefficients, we show that
\begin{equation}
	\left(\begin{array}{c} \hat{P}^{TE}_{mn}
          \\ \hat{P}^{TB}_{mn}
	\end{array}\right)=\displaystyle\sum_{m_1,n_1}\left(\begin{array}{ccc}
		M^{gaid}_{mn,m_1n_1} & M^{ffo}_{mn,m_1n_1} \\
		-M^{ffo}_{mn,m_1n_1} & M^{gaid}_{mn,m_1n_1}
	\end{array}\right)\left(\begin{array}{c}
		{P}^{{TE}}_{m_1n_1} \\
		{P}^{{TB}}_{m_1n_1}
	\end{array}\right),
\end{equation}
with
\begin{eqnarray}
	M^{gaid}_{mn,m_1n_1}&=&\cos{(2\phi_{mn}^{m_1n_1})}|K_{m,m_1}^{n,n_1}|^2, \\
	M^{ffo}_{mn,m_1n_1}&=&\sin{(2\phi_{mn}^{m_1n_1})}|K_{m,m_1}^{n,n_1}|^2.
\end{eqnarray}
As for the $cross$ blocks in the polarization case, the azimuthal average of
$M^{ffo}_{mn,m_1n_1}$ vanishes, i.e. $\int\int M^{ffo}_{mn,m_1n_1}d\theta
d\theta_1=0$. However, before this averaging, it is not {\it a priori} zero.

\end{appendix}

\end{document}